\documentclass[12pt]{article}
\usepackage[hmargin=1.1 in]{geometry}
\usepackage{setspace}
\usepackage[bottom]{footmisc}
\usepackage{indentfirst}
\usepackage{graphicx,xcolor}
\usepackage{rotating}
\usepackage{mathrsfs}
\usepackage{url}
\usepackage[bf]{caption}
\usepackage{float}
\usepackage{amsmath, amsthm, amssymb, amsfonts, amstext, amsbsy}
\usepackage[authoryear,round]{natbib}
\usepackage{ulem}
\usepackage{multirow}
\usepackage{bbm}
\usepackage{subfig}

\theoremstyle{definition}

\numberwithin{equation}{section}
\begin{document}
\bibliographystyle{plainnat}

\title{Randomization Tests for Equality in Dependence Structure}
\author{Juwon Seo \thanks{The author would like to thank the seminar participants at the 1st International Conference on Econometrics and Statistics, 70th European Meeting of the Econometric Society, 2018 Asian Meeting of the Econometric Society, 2018 Australasia Meeting of the Econometric Society, Chinese University of Hong Kong, SZ, Seoul National University, Texas A\&M University, University of Connecticut, Yonsei University and Aureo de Paula, Brendan K. Beare, Zheng Fang for valuable discussions, Shihao Zhou for excellent research assistance, and Bruno R\'{e}millard for kindly sharing the codes. This work was supported by AcRF Tier 1 Grant under GN. $R12200026311$.}
 \bigskip \\ Department of Economics\\ National University of Singapore }

\maketitle


\sloppy%


\begin{center}
\textbf{Abstract}\\
\end{center}
We develop a new statistical procedure to test whether the
dependence structure is identical between two groups. Rather than relying on
a single index such as Pearson's correlation coefficient or Kendall's $\tau$, we consider
the entire dependence structure by investigating the dependence functions (copulas). The critical values are obtained by a modified randomization procedure designed to exploit asymptotic group invariance conditions. Implementation of the test is intuitive and simple, and does not require any
specification of a tuning parameter or weight function. At the same time,
the test exhibits excellent finite sample performance, with
the null rejection rates almost equal to the nominal level even when the
sample size is extremely small. Two empirical applications concerning the dependence between income and consumption, and the Brexit effect on European financial market integration are provided.

\pagebreak

\doublespacing

\section{Introduction}
\cite{}
As the most fundamental measure of dependence, \textit{Pearson's coefficient
of correlation} has been widely used for centuries in numerous empirical
studies concerning the dependence between variables. Despite its lasting
popularity, however, Pearson's coefficient of correlation certainly has some
limitations in characterizing dependence structures because it only captures
pairwise and linear dependence. When the variables of interest are Gaussian
where the linearity is implied, the correlation coefficient serves as the
best measure of dependence in the sense that the dependence structure is
fully described by the correlation. Nevertheless, it is not always
reasonable to presume that the underlying dependence structure is linear and
in fact, many economic and financial data frequently exhibit
nonlinear relationships.

Alternatively, we may aim to provide a meaningful description of the entire
dependence structure rather than summarize it using a
single index, particularly if nonlinear features of the
variables such as asymmetric dependence or tail dependence are of interest.
In this context, \textit{dependence functions}, also known as \textit{copulas}, have proved to be useful in the studies of dependence structures. Suppose that
$X_{1},...,X_{d}$ are random variables with joint distribution function $H$ and
univariate margins $F_{1},...,F_{d}$. Sklar's theorem
(1959) ensures the existence of a joint distribution function $C:[0,1]^{d}\rightarrow \lbrack
0,1]$ that has uniform margins and satisfies
\begin{equation}
H(x_{1},...,x_{d})=C(F_{1}(x_{1}),...,F_{d}(x_{d}))  \tag{1}
\end{equation}%
for all $(x_{1},...,x_{d})\in \mathbb{R}^{d}$. The function $C$ is
called the copula associated with $H$. This result suggests that any joint
distribution can be decomposed into two parts, the univariate marginal distributions which determine the behavior of individual variables, and the copula function which determines the dependence structure between variables. For the reason, copulas have been major concerns in various
applications of dependence modelling.\footnote{For instance, Patton (2006) used
time-varying copulas to capture the asymmetric dependence between exchange rates. Zimmer and Trivedi (2006) employed trivariate copulas for an application to
family health care demand, and Bonhomme and Robin (2009) explored copulas
in the modelling of the\ transitory component in earnings. In finance,
Embrechts et al.\ (2002) and Rosenberg and Schuermann (2006)
studied risk management using the copula models, while Oh and Patton (2013,
2017) used copulas to model the dependence between stock returns.
Copula-based models of serial dependence or heteroskedasticity have been studied by Chen and Fan (2006), Chen et al.\ (2009), Lee and Long (2009), Ibragimov (2009), Smith et al. (2010), Beare (2010), and Beare and Seo (2014, 2015), Loaiza-Maya et al. (2018). See also Creal and Tsay (2015) for an application to panel data.}

In this paper, we use the copulas to propose a statistical procedure for testing homogeneity of dependence structure between different groups.
By comparing copulas, our test detects any arbitrary form of dissimilarity in the dependence structure. The test statistic is constructed based on the $L_{p}$ distance between two empirical copulas and critical values are obtained from a novel randomization procedure. 
Implementation of our test is intuitive and straightforward, and does not
require any specification of a tuning parameter or weight function that
can be arbitrarily chosen by a researcher. At the same time, the test
exhibits substantially excellent performance in finite samples, with
the null rejection rates almost equal to the nominal level even when the
sample size is extremely small.

In developing our methodology, we adapt the permutation method used to test distributional equality. A modification is required because the null set of copula equality is strictly larger than the null set of distributional equality. When the problem is to
test the equality of distributions, classical permutation tests deliver exact size control and those tests are as powerful as
standard parametric tests under general conditions (Hoeffding, 1952).
However, when the null hypothesis to be tested is larger than distributional equality, the usual permutation tests
generally do not control size even when the sample size is large
(Romano, 1990; Chung and Romano, 2013). Therefore, we expect that a naive application of a permutation test to the hypothesis of copula equality may fail to deliver valid inference even asymptotically.

To resolve this problem, recent studies on randomization tests focus on the studentization of a test statistic. See, for instance, Neuhaus (1993), Janssen (1997, 1999, 2005), Janssen and Pauls (2003, 2005), Neubert and Brunner (2007), Omelka and Pauly (2012) and Chung and Romano (2016). These modifications provide asymptotic level $\alpha $ tests of relevant null hypotheses, with exact size control under distributional equality. However, this approach is not applicable in our context because our test statistic has more or less complicated form and its sampling distribution involves several Brownian bridges determined by underlying copulas, and the derivatives of those copulas. Henceforth, we take a different approach introducing theorems of conditional convergence in the context of randomization test literature. Although we only focus on the copula equality, our technical ingredients may be useful for handling other null hypotheses where the test statistic is not linear and the delta method is to be invoked.

The remainder of the paper is structured as follows. In Section 2, we
explain why the classical permutation method is invalid when used to test copula equality. Our main results are
presented in Section 3, where we introduce modified randomization procedures and discuss their asymptotic properties. In Section 4, we report some numerical simulation results. In Section 5, two empirical applications concerning the dependence between income and consumption, and the Brexit effect on European financial market integration are provided. Proofs of lemmas and theorems are collected together in the Appendix.

\section{Application of the permutation method}
Suppose that $X^{1},...,X^{n}$ are i.i.d. draws from a distribution $%
H_{1}\,$and $Y^{1},...,Y^{m}$ are i.i.d. draws from a distribution $H_{2}$. The
two samples are independent. Each distribution is $d$-variate with $d\in\mathbb N$, and we write $X^i=(X_1^i,\ldots,X_d^i)$ and $Y^j=(Y_1^j,\ldots,Y_d^j)$ for $i=1,...,n$ and $j=1,...,m$. We denote the univariate margins of $H_1$ by $F_1,\ldots,F_d$ and the univariate margins of $H_2$ by $G_1,\ldots,G_d$. Let $N$ be the total number of observations (that is, $N=n+m$), and let $W$ be the stacked matrix
of the size $N\times d$,
\begin{equation}
W=(W^{1}; W^{2};...;W^{N})=(X^{1};...;X^{n};Y^{1};...;Y^{m})%
\text{. }  \tag{2}
\end{equation}

For the case of testing the hypothesis $\mathcal H_0:(H_{1},H_{2})\in \Theta_{00} $ with $\Theta_{00} =\{(H,H')\in\mathbb D^2|H=H'\}$, where $\mathbb D$ is the set of all $d$-variate probability distributions, we may easily construct an exact level $\alpha $ test using a randomization test based on permuting the rows of $W$. Complications
arise when the null hypothesis of interest is strictly larger than $%
\Theta_{00} $. If this is the case, it is well known that permutation tests
generally cannot control the probability of Type I error even
asymptotically. Hence, inferences based on a permutation test can be
highly misleading (Romano, 1990; Chung and Romano, 2013).

Let $\Phi :%
\mathbb{D}\rightarrow \ell ^{\infty }([0,1]^{d})$ be the map that
sends a cdf $\tilde{H}\in \mathbb{D}$ with margins $\tilde{F}%
_{1},...,\tilde{F}_{d}$ to $\tilde{H}(\tilde{F}_{1}^{-1},...,\tilde{F}%
_{d}^{-1})$. For each $k=1,...,d$ and $t\in \lbrack 0,1]$, $\tilde{F}%
_{k}^{-1}(t)$ is defined to be $\inf \{x|\tilde{F}_{k}(x)\geq t\}$. In words, given a joint distribution
function, $\Phi $ is the map which provides the corresponding copula as an
output. Now we can formulate our null hypothesis of copula equality as%
\begin{equation}
\mathcal H_0:(H_{1},H_{2})\in \Theta _{0}\text{ with }\Theta
_{0}=\{(H,H')\in\mathbb D^2|\Phi (H)=\Phi (H')\}\text{.}  \tag{3}
\end{equation}%
Since we may have two different distributions $H_{1}$ and $H_{2}$ which
share the same copula, our null set $\Theta _{0}$ is strictly larger than $%
\Theta_{00} $. This implies that a randomization test based on the permutations
of $W$ may possibly lead to a permutation distribution which does not
agree with the correct limit of our test statistic.

To illustrate this point, let's consider the following example with two
bivariate ($d=2$) cdfs $H_{1}$ and $H_{2}$. We let $H_{1}$ be the
distribution of $X_{1}$ and $X_{2}$ where $X_{1}$ is uniformly distributed
between zero and one, and $X_{2}$ is defined to be $X_{1}+1$. Now consider
another distribution $H_{2}$, that is the distribution of $Y_{1}$ and $Y_{2}$%
, where $Y_{2}$ is uniformly distributed between zero and one, and $Y_{1}$
is defined to be $Y_{2}+1$. The joint distributions $H_{1}$ and $H_{2}$
are given by $H_{1}(x_{1},x_{2})=\min (x_{1},x_{2}-1)$ on $(x_{1},x_{2})\in
\lbrack 0,1]\times \lbrack 1,2]$ and $H_{2}(x_{1},x_{2})=$ $\min
(x_{1}-1,x_{2})$ on $(x_{1},x_{2})\in \lbrack 1,2]\times \lbrack 0,1]$,
respectively. See Figure 2.1 for a graphical illustration of the shapes of
the distributions. For both $H_{1}$ and $H_{2}$,\ it is easy to
verify that the associated copula is $C_{1}(u_{1},u_{2})=C_{2}(u_{1},u_{2})=%
\min (u_{1},u_{2})$ for $(u_{1},u_{2})\in \lbrack 0,1]^{2}$. We see that $%
H_{1}$ and $H_{2}$ are different but their copulas $C_{1}$ and $C_{2}$ are
identical. In other words, $(H_{1},H_{2})\in \Theta _{0}$ but $%
(H_{1},H_{2})\notin \Theta_{00} $.

\begin{figure}%
    \centering
    \subfloat[]{{\includegraphics[width=5.7cm]{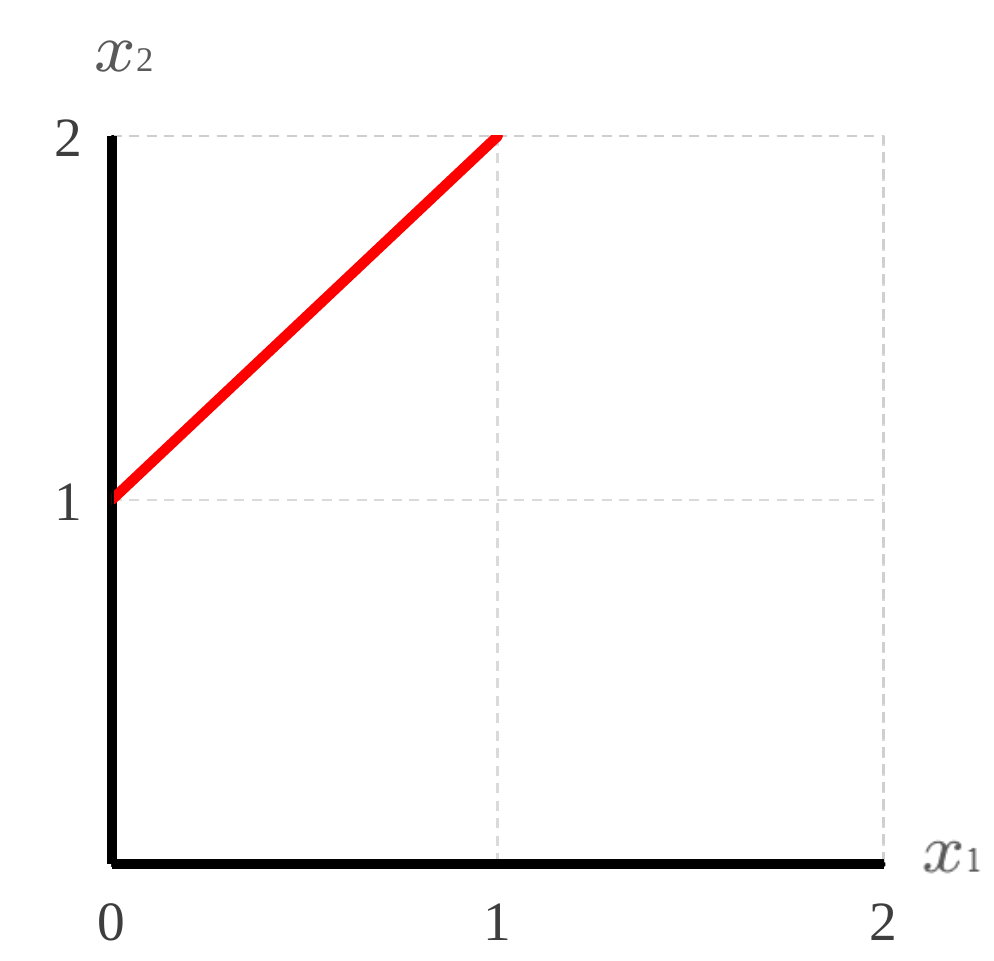} }}%
    \qquad
    \subfloat[]{{\includegraphics[width=5.7cm]{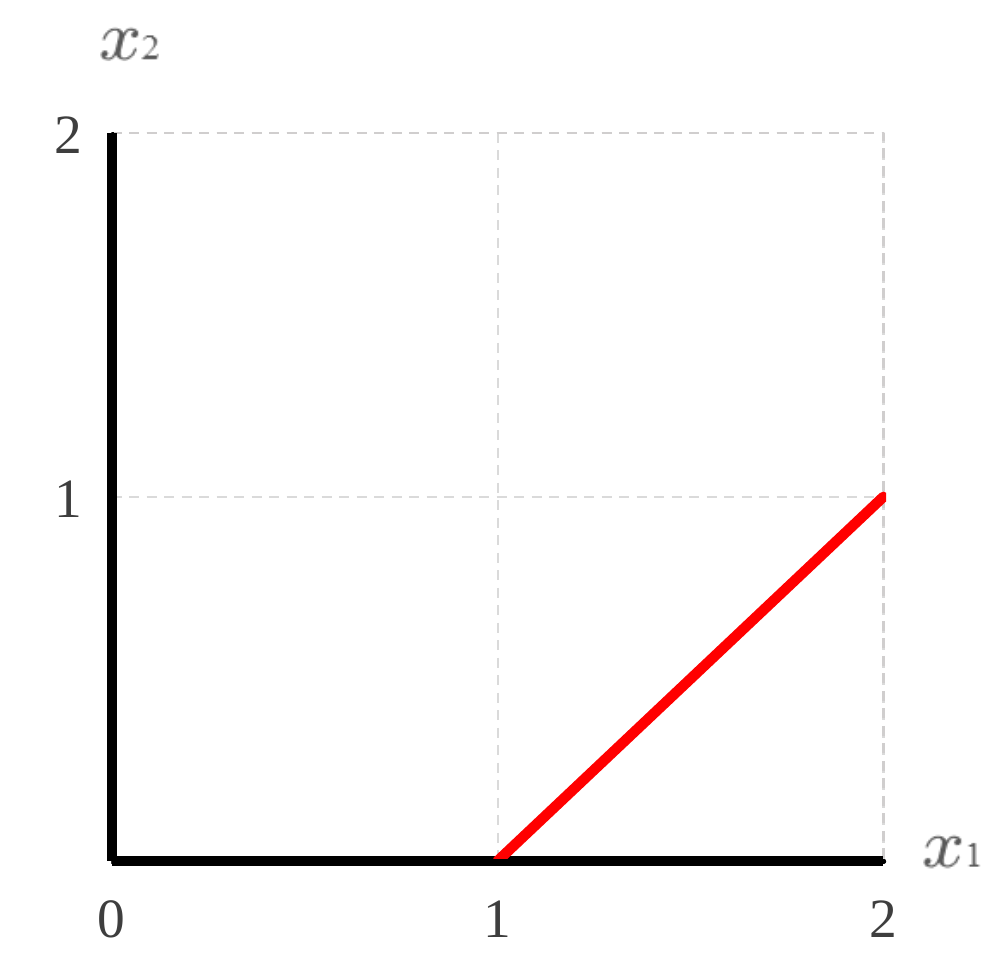} }}%
    \caption{In our example, $X_{1}$ and $X_{2}$ with the joint distribution $H_{1}$ are distributed evenly over the
the bold line displayed in Panel (a) while $Y_{1}$ and $Y_{2}$ with the joint distribution $H_{2}$ are distributed evenly over the bold line displayed in Panel (b).}%
\end{figure}

Suppose we will use a test statistic $T_{n,m}$ to test copula equality. The limit distribution of $T_{n,m}$ may generally
depend on the underlying copulas and we hope to approximate it using the permutation
distribution of $T_{n,m}$. Since the permuted samples behave as though drawn from a mixture of $H_1$ and $H_2$, the permutation distribution of $%
T_{n,m}$ can be inferred from its unconditional distribution when the samples are drawn from that mixture distribution. This
argument is well explained in Chung and Romano (2013), where the authors
employed a contiguity argument and coupling construction assuming the linearity of $T_{n,m}$. Although our test statistic in the next section is not linear with respect to the original sample, the validity of our test procedure can be demonstrated by applying results in Chung and Romano (2013) to a simpler infeasible test statistic whose construction depends on the univariate margins being known. Viewing our test in this light, it is natural to ask how the copula of a mixture of $H_1$ and $H_2$ is determined.

Now return to our example and consider a mixture of $H_{1}$ and
$H_{2}$ defined by $\bar{H}=\lambda H_{1}+(1-\lambda )H_{2}$ with some
weight $\lambda \in \lbrack 0,1]$. The support of the mixture distribution $%
\bar{H}$ is displayed in Panel (a) of Figure 2.2. When $\lambda =1/2$ for
instance, $\bar{H}$ corresponds to the distribution of a pair of random
variables uniformly distributed over the two diagonals in Panel (a) of
Figure 2.2. The corresponding copula in this case is,
\begin{equation*}
C^{\dagger }(u_{1},u_{2})=
\begin{array}{c}
\begin{cases}
\min (u_{1},u_{2}-\frac{1}{2})\text{ \ \ \ \ for }(u_{1},u_{2})\in \left[ 0,%
\frac{1}{2}\right] \times \left[ \frac{1}{2},1\right] \\
\min (u_{1}-\frac{1}{2},u_{2})\text{ \ \ \ \ for }(u_{1},u_{2})\in \left[
\frac{1}{2},1\right] \times \left[ 0,\frac{1}{2}\right] \\
\max (u_{1}+u_{2}-1,0)\text{ \ \ \ \ \ \ otherwise,}%
\end{cases}
\end{array}%
\end{equation*}
which has the uniform probability mass over the bold lines in Panel (b) of
Figure 2.2. We observe from this example that the copula associated with the
mixture distribution $\bar{H}$ can be different from $C_{1}$ (or $C_{2}$)
even when $C_{1}=C_{2}$.

\begin{figure}%
    \centering
    \subfloat[]{{\includegraphics[width=5.7cm]{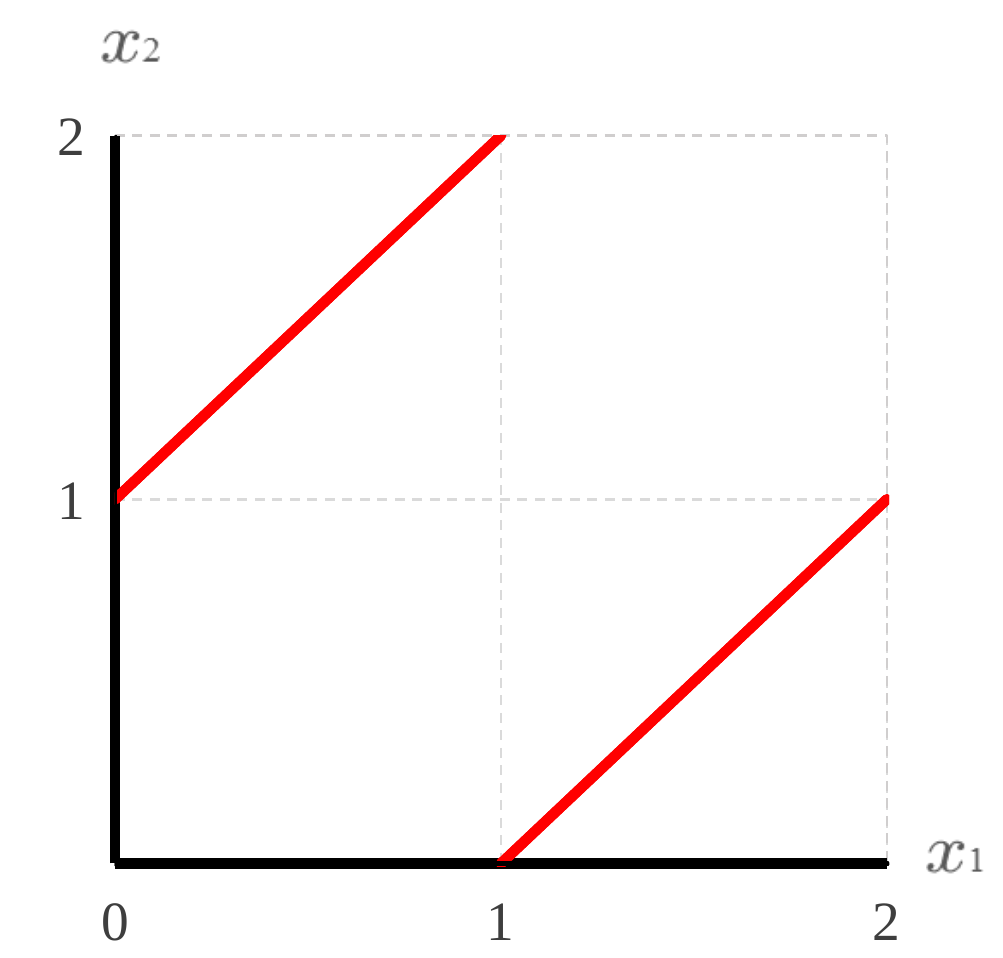} }}%
    \qquad
    \subfloat[]{{\includegraphics[width=5.7cm]{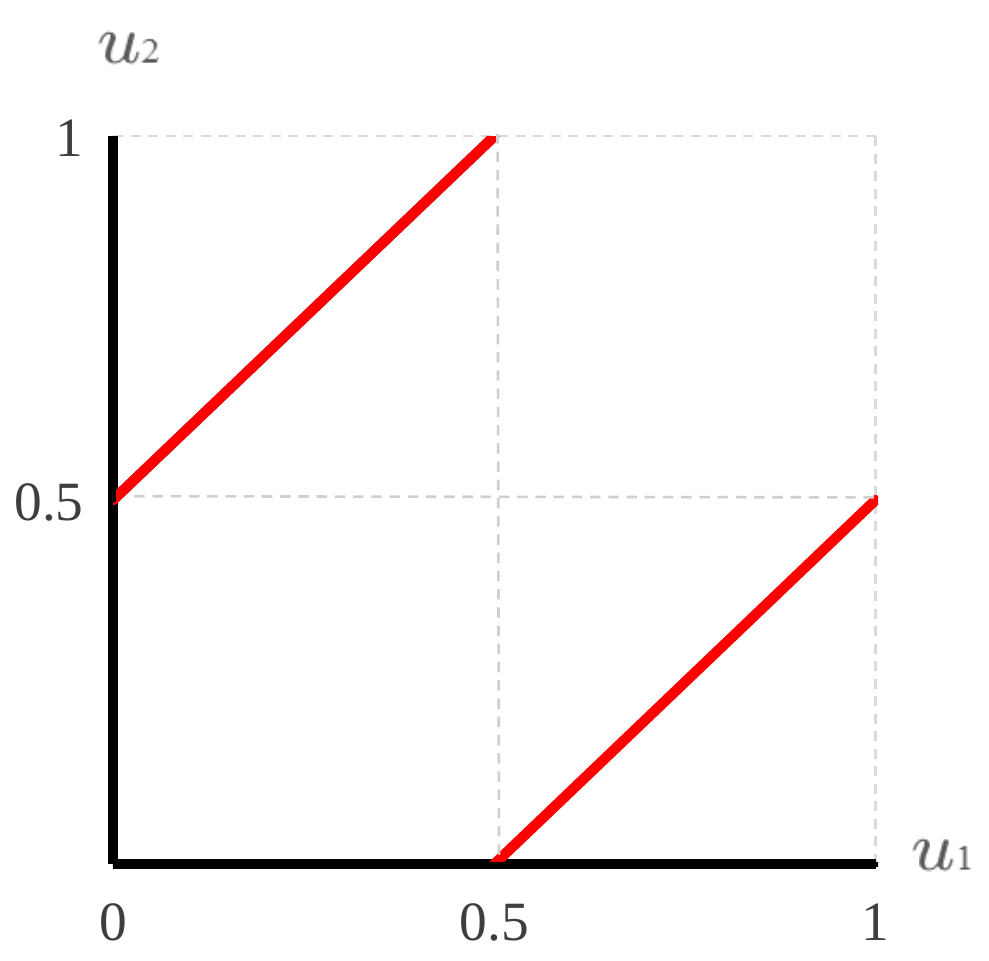} }}%
    \caption{Panel (a) displays the
support of $\bar{H}$ in our example. When $\lambda =1/2$, $\bar{H}$ is the distribution of
a pair of random variables uniformly distributed over the two diagonals.
Panel (b) describes the corresponding copula associated with $\bar{H}$ when $\lambda =1/2$.}%
\end{figure}

The discussion in the preceding paragraph should provide some insight into why a
permutation test obtained by permuting the rows of $W$ can be invalid.
Failure of such a permutation test may be attributed to the fact that (i) the limit
distribution of $T_{n,m}$ is determined by the true underlying copulas in general, and
(ii) unless $\lambda$ is zero or one, the copula of the mixture distribution
is different from $C_{1}$ or $C_{2}$. In our example above, the limit of $%
T_{n,m}$ is determined by $C_{1}$ which is equal to $C_{2}$, whereas
the limit of the permutation distribution is determined by the copula
associated with the mixture distribution with $\lambda$ being determined by the limiting value of $n/(n+m)$. This suggests that estimating the asymptotic null distribution of $T_{n,m}$ by permuting the samples from $H_{1}$ and $H_{2}$ may not be adequate.

A natural solution to this problem is to manipulate the test statistic in
such a way that its limiting distribution does not depend on the underlying
probability distributions or copulas. When applied to a properly studentized test
statistic, a permutation test can deliver asymptotically valid inference
in the sense that the rejection probability converges to the nominal level $\alpha$ when the
null hypothesis is true. However, this can only be done in limited circumstances where the test statistic is of a simple linear form. For more general applications of the permutation method, it requires more technical development.

To adapt the permutation method for the inference of copula equality, we will instead commence from the observation that copulas can be regarded as the joint distribution of the probability integral transform of
each univariate variable. For instance, the copula in (1) is the joint
distribution of $d$ random variables, $F_{1}(X_{1}),...,F_{d}(X_{d})$. From
this point of view, the problem of testing copula equality is closely
related to the problem of testing the equality of probability
distributions; assuming that the univariate margins $F_{1},...,F_{d}$ are
known, the classical theory of randomization tests applies and an exact $%
\alpha $ level test can be constructed by the permutation method. However,
the marginal distributions are not known in practice and can only be estimated
consistently, and this suggests that we may only rely on asymptotic group invariance conditions in our application of the permutation method. In this respect, our results in the next section complement those of Canay et al. (2017) or Beare and Seo (2017), who
investigated the behavior of randomization tests under approximate
symmetry conditions, albeit with different formalizations of approximate symmetry.

\section{Test construction}
In this section, we explain how to construct valid permutation tests of the hypothesis of equal dependence structure. Two asymptotically valid procedures are proposed. The first is similar to the multiplier technique of R\'emilard and
Scaillet (2009). The second is a modification of the first that eliminates the need for partial derivative estimation.
Our main results are summarized in Theorem 3.1 and Theorem 3.2.
Although we confine our attention to the two-sample problem for simplicity,
it is straightforward to extend our results to the more general $k$-sample problem.

We start out by defining the test statistic. As in the previous section, let $C_{1}$ and $%
C_{2}$ be the copulas associated with $H_{1}\,$and $H_{2}$ respectively; that is, $%
C_{1}=\Phi (H_{1})$ and $C_{2}=\Phi (H_{2})$. Since our null hypothesis is
given as in (3), a proper test statistic can be constructed based on the
discrepancy between $C_{1}$ and $C_{2}$. 
Let $\hat{F}_{1,n},...,\hat{F}_{d,n},\hat{G}_{1,m},..,\hat{G}_{d,m}$
be the empirical distributions corresponding to $F_{1},...,F_{d}$, $%
G_{1},...,G_{d}$,
\begin{equation}
\hat{F}_{q,n}(x)=\frac{1}{n}\sum_{i=1}^{n}1(X_{q}^{i}\leq x),\quad \hat{G}%
_{q,m}(x)=\frac{1}{m}\sum_{j=1}^{m}1(Y_{q}^{j}\leq x)\text{ for }q=1,...,d%
\text{ and }x\in \mathbb{R}\text{, \ }  \tag{4}
\end{equation}%
and $\hat{C}_{1,n}$, $\hat{C}_{2,m}$ be the empirical copulas computed from
the two independent samples as below.%
\begin{eqnarray*}
\hat{C}_{1,n}(u_{1},...u_{d}) &=&\frac{1}{n}\sum_{i=1}^{n}1\{ \hat{F}%
_{1,n}(X_{1}^{i})\leq u_{1},...,\hat{F}_{d,n}(X_{d}^{i})\leq u_{d}\}
\text{ and} \\
\hat{C}_{2,m}(u_{1},...u_{d}) &=&\frac{1}{m}\sum_{i=1}^{m}1\{ \hat{G}%
_{1,m}(Y_{1}^{j})\leq u_{1},...,\hat{G}_{d,m}(Y_{d}^{j})\leq u_{d}\}
\text{ for }(u_{1},...u_{d})\in \lbrack 0,1]^{d}\text{.}
\end{eqnarray*}
Our test statistic is provided by
\begin{equation*}
T_{n,m}^{(p)}=\sqrt{\frac{nm}{n+m}}\left\Vert \hat{C}_{1,n}-\hat{C}%
_{2,m}\right\Vert _{p},
\end{equation*}%
where $\Vert \cdot \Vert _{p}$ is the $L_{p}$ norm with respect to the
Lebesgue measure on $[0,1]^{d}$ given $p\in \lbrack 1,\infty ]$.

The empirical copula is the most widely used nonparametric estimator of copulas and its asymptotic results have been well established in the literature. The limit theory of the empirical copula process was firstly developed by Deheuvels (1981a, 1981b) under independence, and extended to nonindependent
cases by Gaenssler and Stute (1987) in the Skorokhod space $D([0,1]^{2})$
and later, by Fermanian et al. (2004) in the space $\ell
^{\infty }\left( [0,1]^{2}\right) $. See also Segers (2012), van der Vaart
and Wellner (1996, 2007) and Tsukahara (2005) for more results. As in those papers, we henceforth assume that $F_{1},...,F_{d},G_{1},...,G_{d}$ are
continuous, and that $C_{1}$ and $C_{2}$ admit continuous partial derivatives on $%
[0,1]^{d}$ to ensure the weak convergence of the empirical copula process. In what follows, let $\rightsquigarrow $ denote Hoffmann-J\o rgensen convergence in \(\ell^\infty([0,1]^d)\) as $\min (n,m)$ tends to infinity.

\bigskip
\textit{\textbf{Lemma 3.1.}} Suppose that $\lambda _{n,m}\equiv n/(n+m)\rightarrow
\lambda $ as $\min (n,m)\rightarrow \infty $. Then we have%
\begin{equation*}
\sqrt{\frac{nm}{n+m}}\{(\hat{C}_{1,n}-\hat{C}_{2,m})-(C_{1}-C_{2})\}%
\rightsquigarrow \sqrt{1-\lambda }\mathbb{C}_{C_{1}}-\sqrt{\lambda }\mathbb{C%
}_{C_{2}},
\end{equation*}%
where $\mathbb{C}_{C_{1}}$ and $\mathbb{C}_{C_{2}}$ are the weak limits of
the empirical copula processes $\sqrt{n}(\hat{C}_{1,n}-C_{1})$ and $\sqrt{m}(%
\hat{C}_{2,m}-C_{2})$ respectively. Accordingly, when $C_1$ and $C_2$ are identical,
\begin{equation*}
\hat{T}_{n,m} \equiv  \sqrt{\frac{nm}{n+m}}(\hat{C}_{1,n}-\hat{C}_{2,m})\rightsquigarrow \sqrt{%
1-\lambda }\mathbb{C}_{C_{1}}-\sqrt{\lambda }\mathbb{C}_{C_{2}}\text{.}
\end{equation*}

The specific forms $\mathbb{C}_{C_{1}}$ and $\mathbb{C}_{C_{2}}$ are determined by the $C_{1}$-Brownian bridge, $C_{2}$-Brownian
bridge, and the partial derivatives of $C_{1}$ and $C_{2}$. Let $\mathbf{u}$ denote the vector of $d$ entries, $(u_{1},...,u_{d})\in
\lbrack 0,1]^{d}$ and $\mathbf{u}_{(q)}=(1,1...,u_{q},...,1,1)\in \lbrack
0,1]^{d}$ denote the vector which has $u_{q}$ as its $q$-th entry and $1$
elsewhere. For a copula $C$, define $\mathbb{B}_{C}$ to be a Brownian bridge
on $[0,1]^{d}$ with covariance kernel
\begin{equation}
\text{Cov}\left( \mathbb{B}_{C}(\mathbf{u}),\mathbb{B}_{C}(\mathbf{u}%
^{\prime })\right) =C(\mathbf{u}\wedge \mathbf{u}^{\prime })-C(%
\mathbf{u})C(\mathbf{u}^{\prime })\text{, \ }  \tag{5}
\end{equation}
where $\mathbf{u}\wedge \mathbf{u}^{\prime }$ be the minimum taken
componentwise, i.e., $(\mathbf{u}\wedge \mathbf{u}^{\prime })=(u_{1}\wedge
u_{1}^{\prime },...,u_{d}\wedge u_{d}^{\prime })$. Then, $\mathbb{C}_{C_{1}}$
and $\mathbb{C}_{C_{2}}$ can be expressed as
\begin{equation}
\mathbb{C}_{C_{l}}(\mathbf{u})=\mathbb{B}_{C_{l}}(\mathbf{u}%
)-\sum\limits_{q=1}^{d}\partial _{q}C_{l}(\mathbf{u})\mathbb{B}_{C_{l}}(%
\mathbf{u}_{(q)})\text{\ \ \ \ for }l=1,2\text{, }  \tag{6}
\end{equation}
with $\partial _{q}C_{l}(\mathbf{u})$ denoting the partial derivative of $%
C_{l}(\mathbf{u})$ with respect to the $q$-th argument, for $q=1,...,d$
and $l=1,2$. The asymptotic null distribution of our test statistic $T_{n,m}^{(p)}$ can be obtained by applying the continuous mapping theorem to the weak convergence established in Lemma 3.1 as,
\begin{equation}
T_{n,m}^{(p)}\rightsquigarrow \mathbb{T}^{(p)}\equiv \left\Vert \sqrt{1-\lambda }\mathbb{C}%
_{C_{1}}-\sqrt{\lambda }\mathbb{C}_{C_{2}}\right\Vert _{p}\text{ .\ \ }
\tag{7}
\end{equation}%

Now we seek to use the permutation method to approximate the limit of our test statistic in (7), under copula equality. Before entering into the main
analysis, it is worth noting that when the univariate margins are known, an
exact level $\alpha$ test can be delivered by the permutation method. This
is because $C_{1}$ is the joint distribution of $%
F_{1}(X_{1}),...,F_{d}(X_{d})$ and $C_{2}$ is the joint distribution of $%
G_{1}(Y_{1}),...,G_{d}(Y_{d})$. Therefore, given the univariate margins $%
F_{1},...,F_{d}$ and $G_{1},...,G_{d}$, copula equality can be
reformulated as distributional equality.\footnote{%
See Remark 3.2 for conditions under which copula equality is necessary and
sufficient for distributional equality.} In the application of the
permutation method, however, permutations should be applied to the transformed data
\begin{equation}
Z=(Z^{1};Z^{2};...;Z^{N})=(U^{1};...;U^{n};V^{1};...;V^{m})  \tag{8}
\end{equation}%
as opposed to (2), properly accounting for the group invariance conditions. Here, the vectors $U^{i}=(F_{1}(X_{1}^{i}),...,F_{d}(X_{d}^{i}))$ and $%
V^{j}=(G_{1}(Y_{1}^{j}),...,G_{d}(Y_{d}^{j}))$ are the probability
integral transforms of the $i$-th and $j$-th observations of $X$ and $Y$
respectively, for $i=1,...,n$ and $j=1,...,m$.

When the univariate margins are known, $C_{1}$ and $C_{2}$ can be estimated by the empirical
distributions computed from $\{U^{i}\}_{i=1}^{n}$ and $\{V^{j}\}_{j=1}^{m}$, respectively. Let $\tilde{C}_{1,n}(Z)(\mathbf{u})$ $=$ $\frac{1}{n}%
\sum_{i=1}^{n}1(U^{i}\leq \mathbf{u})$ and $\tilde{C}_{2,m}(Z)(\mathbf{u})=%
\frac{1}{m}\sum_{j=1}^{m}1(V^{j}\leq \mathbf{u})$ be the empirical
distributions, and let $\tilde{T}_{n,m}(Z)$ be a scaled difference
between the two empirical distributions, i.e.,
\begin{equation*}
\tilde{T}_{n,m}(Z) \equiv  \sqrt{\frac{nm}{n+m}}(\tilde{C}_{1,n}(Z)-\tilde{C}_{2,m}(Z))%
\text{.}
\end{equation*}%
The permutation distribution of $\tilde{T}_{n,m}$ can be found by verifying the Hoeffding's condition (Hoeffding, 1952). This is done in the next Lemma 3.2, where the joint convergence in (9) implies that the permutation distribution of $\tilde{T}_{n,m}$ converges to $\mathbb{\tilde{T}}$ defined therein, as $n$ and $m$ tend to infinity. In Lemma A.1 in Appendix, we additionally show that this limit process $\mathbb{\tilde{T}}$ coincides with the limit of $\tilde{T}_{n,m}(Z)$ under the null hypothesis. As a direct consequence, the permutation test based on $\Vert \tilde{T}_{n,m}(Z)\Vert _{p}$, which can be regarded as a version of our test statistic $T_{n,m}^{(p)}$ for known margins, controls the size of the test asymptotically.\footnote{%
We can also verify that this test is exact employing the usual proof techniques
for the permutation method (Lemma 3.2 is only an asymptotic result).}

In the following, let $\mathbf{G}_{N}$ be the set of all permutations of $%
\{1,...,N\}$ and $Z^{\pi }$ $=(Z^{\pi (1)};...;Z^{\pi (N)})$ be the permuted
sample for a permutation $\pi =(\pi (1),\pi (2),...,\pi (N))$ drawn from the
uniform distribution on $\mathbf{G}_{N}$ independently of the data. We
further denote $\pi ^{\prime }$ to be a permutation drawn from the uniform
distribution on $\mathbf{G}_{N}$ independently of $\pi $ and the data.

\bigskip
\textit{\textbf{Lemma 3.2.}} Under the assumption that $\lambda _{n,m}\equiv
n/(n+m)=\lambda +O((n+m)^{-1/2})$, we have the convergence
\begin{equation}
(\tilde{T}_{n,m}(Z^{\pi }),\tilde{T}_{n,m}(Z^{\pi ^{\prime
}}))\rightsquigarrow (\mathbb{\tilde{T}},\mathbb{\tilde{T}}^{\prime }),   \tag{9}
\end{equation}%
where $\mathbb{\tilde{T}}$ is a Gaussian process that can be written as
\begin{equation*}
\mathbb{\tilde{T}}\equiv \sqrt{1-\lambda }\mathbb{B}_{\bar{C}}-\sqrt{\lambda
}\mathbb{B}_{\bar{C}}^{\prime } \text{ \ \ \ \ with \ } \bar{C}=\lambda C_{1}+(1-\lambda )C_{2}.
\end{equation*}%
Here, $\mathbb{\tilde{T}}%
^{\prime }$ is an independent copy of $\mathbb{\tilde{T}}$, and $\mathbb{B}_{%
\bar{C}}^{\prime }$ is an independent copy of $\mathbb{B}_{\bar{C}}$.
\bigskip

In applying the result in Lemma 3.2, however, we may encounter a practical problem because the construction of $Z$ in (8) is actually
not feasible when the marginal distributions $F_{1},...,F_{d}$ and $G_{1},...,G_{d}$ are unknown. In the next step, we
shall drop the condition that the univariate margins are known and
instead, consider the permutations based on $\hat{Z}$,%
\begin{equation}
\hat{Z}=(\hat{Z}^{1};...;\hat{Z}^{N})=(\hat{U}%
_{n}^{1};...;\hat{U}_{n}^{n};\hat{V}_{m}^{1};...;\hat{V}_{m}^{m})\text{ }
\tag{10}
\end{equation}%
in which we estimate each univariate margin using its empirical
distribution. Hence now, for $i=1,...,n$ and $j=1,...,m$, we have%
\begin{equation*}
\hat{U}_{n}^{i}=(\hat{U}_{1,n}^{i},...,\hat{U}_{d,n}^{i})\text{ and }\hat{V}%
_{m}^{j}=(\hat{V}_{1,m}^{j},...,\hat{V}_{d,m}^{j})
\end{equation*}%
where $\hat{U}_{q,n}^{i}=\hat{F}_{q,n}(X_{q}^{i})$ and $\hat{V}_{q,m}^{j}=%
\hat{G}_{q,m}(Y_{q}^{j})$ for each\ $q=1,...,d$. Note that under this specification, $\hat{C}_{1,n}$ and $\hat{C}_{2,m}$ can be written by
\begin{equation*}
\hat{C}_{1,n}(\mathbf{u})=\frac{1}{n}\sum_{i=1}^{n}1(\hat{Z}^{i}\leq \mathbf{u})\text{ and }\hat{C}_{2,m}(\mathbf{u})=\frac{1}{m%
}\sum_{i=n+1}^{N}1(\hat{Z}^{i}\leq \mathbf{u}).
\end{equation*}%
Also, $\hat{T}_{n,m}$ in Lemma 3.1 is equivalent to $\tilde{T}_{n,m}(\hat{Z})$.

\bigskip
\textit{\textbf{Lemma 3.3.}} For $\pi\in \mathbf{G}_{N}$, let $\hat{Z}^{\pi }=(\hat{Z}^{\pi (1)};...;\hat{Z}^{\pi (N)})$ be the permuted sample of $\hat{Z}$. Define $\tilde{C}_{1,n}^{\pi }$ and $\tilde{C}%
_{2,m}^{\pi }$ by
\begin{equation*}
\tilde{C}_{1,n}^{\pi }(\mathbf{u})=\frac{1}{n}\sum_{i=1}^{n}1(\hat{Z}^{\pi
(i)}\leq \mathbf{u})\text{ and }\tilde{C}_{2,m}^{\pi }(\mathbf{u})=\frac{1}{m%
}\sum_{i=n+1}^{N}1(\hat{Z}^{\pi (i)}\leq \mathbf{u})
\end{equation*}%
so that
$\tilde{C}_{1,n}^{\pi }$ and $\tilde{C}_{2,m}^{\pi }$ are shorthand for $\tilde{C}_{1,n}(\hat{Z}^{\pi})$ and $\tilde{C}_{2,m}(\hat{Z}^{\pi})$ resepectively, and
\begin{equation*}
\tilde{T}_{n,m}(\hat{Z}^{\pi })=\sqrt{\frac{nm}{n+m}}( \tilde{C}%
_{1,n}^{\pi }-\tilde{C}_{2,m}^{\pi }) \text{.}
\end{equation*}%
Under the assumption that $\lambda _{n,m}\equiv n/(n+m)=\lambda +O((n+m)^{-1/2})$, we have
\begin{equation}
(\tilde{T}_{n,m}(\hat{Z}^{\pi }),\tilde{T}_{n,m}(\hat{Z}^{\pi ^{\prime
}}))\rightsquigarrow (\mathbb{\tilde{T}},\mathbb{\tilde{T}}^{\prime }),  \tag{11}
\end{equation}%
where $\mathbb{\tilde{T}}$ and $\mathbb{\tilde{T}}^{\prime }$ are the Gaussian
processes defined in Lemma 3.2.
\bigskip

Lemma 3.3 provides some important implications for the application of permutation method. Firstly, we find that the permutation distribution of $\tilde{T}_{n,m}$ based on the permutations of $\hat{Z}$ is asymptotically equivalent to the one based on the permutations of $Z$, suggesting that $\hat{Z}$ serves as a good proxy for $Z$. However, since the limit of $\tilde{T}_{n,m}(\hat{Z})$ (equivalently, the limit of $\hat{T}_{n,m}$ in Lemma 3.1) is different from the limit of $\tilde{T}_{n,m}(\hat{Z}^{\pi})$ in Lemma 3.3, we can not validly employ the standard randomization procedure to test copula equality based on $\tilde{T}_{n,m}$. 

One solution to this problem is to combine the result in Lemma 3.3 with a proper estimation procedure for the copula derivatives. To understand how, note that when $C_1$ and $C_2$ are identical, $\mathbb{\tilde{T}}$ is equivalent to $\mathbb{\tilde{T}}_{0}\equiv \sqrt{1-\lambda }\mathbb{B}%
_{C_{1}}-\sqrt{\lambda }\mathbb{B}_{C_{2}}$ (see Lemma A.1 in Appendix) which appears in the limit of our test statistic under the null hypothesis. After approximating this term by simulating $%
\tilde{T}_{n,m}(\hat{Z}^{\pi })$ over $\pi \in \mathbf{G}_{N}$, the remaining terms are the partial derivatives of copulas that can be consistently estimated by either smoothed or
non-smoothed versions of estimators such as those in Scaillet (2005), R\'{e}millard and Scaillet (2009), or Segers (2012). We formally state this result as in Theorem 3.1.

\bigskip \textit{\textbf{Theorem 3.1}}. For $\pi\in \mathbf{G}_{N}$, define $%
\mathfrak{T}_{n,m}^{\pi }$ as%
\begin{equation*}
\mathfrak{T}_{n,m}^{\pi }(\mathbf{u})=\tilde{T}_{n,m}(\hat{Z}^{\pi })(%
\mathbf{u})-\sum\limits_{q=1}^{d}\tilde{T}_{n,m}(\hat{Z}^{\pi }(\mathbf{u}%
_{(q)}))\widehat{\partial _{q}C}(\mathbf{u})\text{,}
\end{equation*}%
where $\widehat{\partial _{q}C}$ is a consistent estimator of $\partial _{q}C$ for $q=1,...,d$. When $C_1$ and $C_2$ are identical, we have
\begin{equation*}
P(\left\Vert \mathfrak{T}_{n,m}^{\pi }\right\Vert _{p}\leq c|\hat{Z})%
\xrightarrow{p}P(\mathbb{T}^{(p)}\leq c)
\end{equation*}%
for any continuity points $c\in (0,\infty )$.

In the related work, R\'emilard and Scaillet (2009) also propose a statistical procedure for testing copula equality. Their test statistic is based on the Cram\'{e}r-von Mises distance and the critical values are obtained through an
approximation of each term appearing in (6). More specifically, $\mathbb{B}%
_{C_{l}}$ for $l=1,2$ is approximated by a multiplier bootstrap as in
Scaillet (2005), and the derivatives are estimated individually. Although we may develop a valid permutation test scheme based
on Theorem 3.1, the practical
advantage is relatively small because we simply end up with replacing the multiplier bootstrap part by the randomization procedure. A shortcoming of such tests is that, as
the dimension $d$ increases, the number of derivative terms to be estimated
increases and hence, the test procedure becomes more
complicated and less precise. The problem is aggravated in the $k$-sample problem as more copulas are involved.\footnote{Besides the multiplier bootstrap method in R\'{e}millard and Scaillet (2009), other bootstrap schemes may possibly be applicable to test copula equality. For instance, Fermanian et al.\ (2004) adopted the usual i.i.d. bootstrap based on the sampling with replacement while B\"{u}cher and Dette (2010) proposed a direct multiplier bootstrap which does not require the estimation procedure of the copula derivatives. However, we shall not dwell on them in this paper because these approximations have shown be less accurate in many simulation experiments than the bootstrap method of R\'{e}millard and Scaillet (2009) that involves estimation procedure of copula derivatives. For the reason, recent studies which use bootstrap procedures for the empirical copula process rely more on the multiplier bootstrap with derivative estimation, or its extension. See Kojadinovic and Yan (2011), Kojadinovic et al.\ (2011), Genest et al.\ (2012), and Genest and Ne\v{s}lehov\'{a} (2014).}

We will thus propose another method for approximating the asymptotic null distribution of our test statistic which, unlike the proposal in Theorem 3.1 or the test in R\'{e}millard and Scaillet (2009), does not require an estimation procedure for the copula
derivatives. 
For technical purposes we first introduce the notion of the conditional weak convergence. Let $\mathbf{A}$ be a metric space
and $\xi _{N}^{\pi }\in \mathbf{A}$ be an element which depends on the data
and $\pi $, regarding $\pi $ as a random element uniformly
distributed on $\mathbf{G}_{N}$. Definition 3.1 on the conditional
weak convergence is adopted from Kosorok (2008).

\bigskip
\textit{\textbf{Definition 3.1}}. Let $BL_{1}(\mathbf{A})$ be the set of real
Lipschitz functions on $\mathbf{A}$ that are uniformly bounded by one with Lipschitz constant
bounded by one, let $E_{\pi }$ be the expectation over $\pi $ holding the data
fixed, and let $f(\xi _{N}^{\pi })^{\ast }$ and $f(\xi _{N}^{\pi })_{\ast }$ be
the minimal measurable majorant and maximal measurable minorant of $f(\xi
_{N}^{\pi })$ with respect to the data and random index jointly. Then we have $\xi _{N}^{\pi }\overset{\mathrm{P}}{\underset{\pi }{%
\rightsquigarrow }}\xi $ if
(i) $\sup_{f\in \mathrm{BL}_{1}(\mathbf{A})}|E_{\pi }f(\xi _{N}^{\pi
})-Ef(\xi )|\rightarrow 0$ in outer probability and
(ii) $E_{\pi }f(\xi _{N}^{\pi })^{\ast }-E_{\pi }f(\xi _{N}^{\pi })_{\ast
}\rightarrow 0$ in probability for every $f\in BL_{1}(\mathbf{A})$.
\bigskip

The conditional convergence in Definition 3.1 is frequently employed to verify the validity of bootstrap techniques, and can be also
used to verify the asymptotic validity of other resampling procedures. In our context, the Hoeffding's condition $(\xi _{N}^{\pi },\xi _{N}^{\pi^{\prime }})\rightsquigarrow (\xi ,\xi ^{\prime })$, for a random element $\xi _{N}^{\pi } \in \mathbf{A}$, can be replaced by the conditional convergence, $\xi
_{N}^{\pi }\overset{\mathrm{P}}{\underset{\pi }{\rightsquigarrow }}\xi $. This property is also used in Beare and Seo (2017) and it turned out to be particularly useful when the test statistic of interest has a nonlinear form and the continuous mapping theorem and functional delta method for the conditional convergence are to be invoked. We provide the details of the continuous mapping theorem and the
functional delta method for the conditional convergence in Lemma A.2
and Lemma A.3 in the Appendix.  

Secondly, we also need to introduce a differentiability result of the operator $\Phi $ defined in Section 2. Let $\mathbb{C}$ be the
space of continuous functions on $[0,1]^{d}$ and let $\mathbb{H}$ be the set
of functions $f\in\mathbb C$ which are grounded and pinned at the end point $%
f(1,1,...,1)=0$. Then, by  B\"ucher and Volgushev (2013), $\Phi $ is Hadamard differentiable at any regular copula $C$ in $\mathbb{D}$ tangentially to $\mathbb{H}$, with the
derivative given by%
\begin{equation*}
\Phi _{C}^{\prime }f(\mathbf{u})=f(\mathbf{u})-\sum_{q=1}^{d}\partial _{q}C(%
\mathbf{u})f(\mathbf{u}_{(q)})\text{.}
\end{equation*}%
This expression appears in our next Lemma 3.4, where we apply the functional delta method to the conditional convergence of $\tilde{T}_{n,m}(\hat{Z}^{\pi })$ implied by Lemma 3.3. Since $\bar{C}=\lambda C_{1}+(1-\lambda )C_{2}$ is a regular copula, we have%
\begin{equation}
\Phi _{\lambda C_{1}+(1-\lambda )C_{2}}^{\prime }f(\mathbf{u})=f(\mathbf{u}%
)-\sum_{q=1}^{d}\{\lambda \partial _{q}C_{1}(\mathbf{u})+(1-\lambda
)\partial _{q}C_{2}(\mathbf{u})\}f(\mathbf{u}_{(q)})\text{ }  \tag{12}
\end{equation}%
for any $f\in \mathbb{H}$. Based on this expression, one can easily see that the limit in Lemma 3.4 is finally equivalent to the limit of $\hat{T}_{n,m}$ when the null hypothesis is true.

\bigskip
\textit{\textbf{Lemma 3.4.}} Under the assumption that $\lambda _{n,m}\equiv n/(n+m)=\lambda+o((\min (n,m))^{-1/2})$, we have
\begin{equation*}
\sqrt{\frac{nm}{n+m}}\{ \Phi ( \tilde{C}_{1,n}^{\pi }) -\Phi
( \tilde{C}_{2,m}^{\pi }) \} \overset{\mathrm{P}}{\underset{%
\pi }{\rightsquigarrow }}\Phi _{\lambda C_{1}+(1-\lambda )C_{2}}^{\prime }%
\mathbb{\tilde{T}}
\end{equation*}%
where $\Phi _{\lambda C_{1}+(1-\lambda )C_{2}}^{\prime }\mathbb{\tilde{T}}$
is the Hadamard derivative of $\Phi $ at $\lambda C_{1}+(1-\lambda )C_{2}$
in direction $\mathbb{\tilde{T}}$.
In addition, we have
\begin{equation*}
\sqrt{\frac{nm}{n+m}}\{ \Phi ( \tilde{C}_{1,n}^{\pi }) -\Phi
( \tilde{C}_{2,m}^{\pi }) \} \overset{\mathrm{P}}{\underset{%
\pi }{\rightsquigarrow }}\sqrt{1-\lambda }\mathbb{C}_{C_{1}}-\sqrt{\lambda }\mathbb{C}_{C_{2}},
\end{equation*}%
when $C_{1}$ and $C_{2}$ are identical.
\bigskip

Hence we conclude that, to approximate the correct limit of our test statistic, the permutation distribution should be obtained through computing the process defined in Lemma 3.4.
However, it may not be clear how to compute $\Phi ( \tilde{C}_{1,n}^{\pi }) $ and $\Phi ( \tilde{C}_{2,m}^{\pi }) $ practically. For the last step, our Theorem 3.2  facilitates the computation by showing that we can replace $\Phi ( \tilde{C}_{1,n}^{\pi }) $ and $\Phi ( \tilde{C}_{2,m}^{\pi }) $ with the two empirical copulas computed from the permuted sample of $\hat{Z}$, as the difference is asymptotically negligible. 

\bigskip
\textit{\textbf{Theorem 3.2.}} For $\pi\in \mathbf{G}_{N}$, define $\mathfrak{R}_{n,m}^{\pi }$ to be
\begin{equation*}
\mathfrak{R}_{n,m}^{\pi }=\sqrt{\frac{nm}{n+m}}( \hat{C}_{1,n}^{\pi }-%
\hat{C}_{2,m}^{\pi })
\end{equation*}%
where $\hat{C}_{1,n}^{\pi }$ and $\hat{C}_{2,m}^{\pi }$ are the empirical copulas\footnote{To be explicit, the empirical copulas computed from the samples $\{\hat{Z}^{\pi (i)}\}_{i=1}^{n}$ and $\{\hat{Z}^{\pi (i)}\}_{i=n+1}^{N}$ are,
\begin{eqnarray*}
\hat{C}_{1,n}^{\pi }(\mathbf{u}) &=&\frac{1}{n}\sum_{i=1}^{n}1\left(\mathcal{\hat{%
F}}_{1,n}^{\pi }(\hat{Z}_{1}^{\pi (i)})\leq u_{1},...,\mathcal{\hat{F}}%
_{d,n}^{\pi }(\hat{Z}_{d}^{\pi (i)})\leq u_{d}\right)\text{ and} \\
\text{ }\hat{C}_{2,m}^{\pi }(\mathbf{u}) &=&\frac{1}{m}\sum_{i=n+1}^{N}1\left((\mathcal{\hat{G}}_{1,m}^{\pi }(\hat{Z}_{1}^{\pi (i)})\leq u_{1},...,%
\mathcal{\hat{G}}_{d,m}^{\pi }(\hat{Z}_{d}^{\pi (i)})\leq u_{d}\right)
\end{eqnarray*}
respectively, where $\hat{Z}_{q}^{\pi
(i)}$ is the $q$-th component of $\hat{Z}^{\pi (i)}$, i.e.,  $\hat{Z}^{\pi
(i)}=(\hat{Z}_{1}^{\pi (i)},...,\hat{Z}_{d}^{\pi (i)})$ and
\begin{equation*}
\mathcal{\hat{F}}_{q,n}^{\pi }(x)=\frac{1}{n}\sum_{i=1}^{n}1(\hat{Z}_{q}^{\pi
(i)}\leq x)\text{, }\mathcal{\hat{G}}_{q,m}^{\pi }(x)=\frac{1}{m}%
\sum_{i=n+1}^{N}1(\hat{Z}_{q}^{\pi (i)}\leq x),
\end{equation*}
for $q=1,...,d$ and $x\in \mathbb{R}$.}
computed from $\hat{Z}^{\pi}$, and $c_{\alpha }^{(p)}$ to be the empirical $(1-\alpha )$ quantile of its $L_p$ norm
over $\pi \in \mathbf{G}_{N}$.
Assuming that $\lambda _{n,m}\equiv n/(n+m)=\lambda +o((\min (n,m))^{-1/2})$,
the following statements are true.
\begin{enumerate}
\item[(i)]
When $C_{1}=C_{2}$, $P(T_{n,m}^{(p)}>c_{\alpha }^{(p)}|\hat{Z})\xrightarrow{p}\alpha
$ as $\min (n,m)\rightarrow \infty $.
\item[(ii)] When $C_{1}\neq
C_{2}$, $P(T_{n,m}^{(p)}>c|\hat{Z})\xrightarrow{p}1$ as $\min (n,m)\rightarrow
\infty $, for any constant $c\in (0,\infty )$.
\end{enumerate}
\bigskip

We close this section by providing a step-by-step guideline to implement the test suggested by Theorem 3.2. Our strategy is simple. First,
compute the test statistic $T_{n,m}^{(p)}=\Vert\tilde{T}_{n,m}(\hat{Z})\Vert _{p}=\Vert \sqrt{\frac{nm}{n+m}}( \hat{C}%
_{1,n}-\hat{C}_{2,m}) \Vert _{p}$. Second, we recompute $\Vert \mathfrak{R}_{n,m}^\pi\Vert _{p}=\Vert \sqrt{\frac{nm%
}{n+m}}( \hat{C}_{1,n}^{\pi }-\hat{C}_{2,m}^{\pi }) \Vert _{p}$
for all permutations $\pi \in \mathbf{G}_{N}$ of $\hat{Z}$, and let
their ordered values be%
\begin{equation*}
\Vert \mathfrak{R}_{n,m}\Vert _{p}^{(1)}\leq \Vert \mathfrak{R}_{n,m}\Vert
_{p}^{(2)}\leq ...\leq \Vert \mathfrak{R}_{n,m}\Vert _{p}^{(N!)}\text{.}
\end{equation*}%
For a nominal level $\alpha $, let $h=N!-[\alpha N!]$ where $[\alpha N!]$
denotes the largest integer less than or equal to $\alpha N!$ and let the $h$%
-th largest value of the permutation statistic be $c_{\alpha }^{(p)}$, that is, $c_{\alpha
}^{(p)}=\Vert \mathfrak{R}_{n,m}\Vert _{p}^{(h)}$. Then inference is made
through the randomization test function $\phi^{(p)} (\hat{Z})$ constructed as
\begin{equation*}
\phi^{(p)}(\hat{Z})=\left\{
\begin{array}{c}
1\text{ \ \ when }T_{n,m}^{(p)}>c_{\alpha }^{(p)} \\
a(\hat{Z})\text{ \ when }T_{n,m}^{(p)}=c_{\alpha }^{(p)} \\
0\text{ \ when }T_{n,m}^{(p)}<c_{\alpha }^{(p)}.\tag{13}%
\end{array}%
\right.
\end{equation*}%
Here, $a(\hat{Z})$ is defined by
\begin{equation*}
a(\hat{Z})=\frac{\alpha n!-M^{+}(\hat{Z})}{M^{0}(\hat{Z})}
\end{equation*}%
where $M^{+}(\hat{Z})$ and $M^{0}(\hat{Z})$ are the number of values of $%
\left\Vert \mathfrak{R}_{n,m}\right\Vert _{p}^{(i)}$ which are greater than $%
\left\Vert \mathfrak{R}_{n,m}\right\Vert _{p}^{(h)}$ and equal to $%
\left\Vert \mathfrak{R}_{n,m}\right\Vert _{p}^{(h)}$, respectively. By Theorem 3.2 (i), this procedure delivers a test with limiting rejection rate equal to nominal size whenever the null hypothesis is true. Theorem 3.2 (ii) establishes the consistency of the test procedure.

In the second stage of implementation, one should be cautious not to compute $\Vert\tilde{T}_{n,m}(\hat{Z}^{\pi})\Vert _{p}$ over $\pi \in \mathbf{G}_{N}$. Since we have $\tilde{T}_{n,m}(\hat{Z})=\sqrt{\frac{nm}{n+m}}( \hat{C}_{1,n}-\hat{C}_{2,m})$, the permutation statistic $\tilde{T}_{n,m}(\hat{Z}^{\pi})$ can be easily employed to approximate the limit of $\sqrt{\frac{nm}{n+m}}( \hat{C}_{1,n}-\hat{C}_{2,m})$ without much caution. However, this is not correct as we discussed in Lemma 3.3. Given the data, the empirical copula for each group is equivalent to the empirical distribution computed from $\hat{Z}$, while the same argument is not true with regard to the permuted samples. Due to the distortion that permutation causes, each univariate component of $\hat{Z}$ no longer approximates a uniform random variable in each group after the permutations, and this leads to a discrepancy between $\tilde{T}_{n,m}(\hat{Z}^{\pi})$ and $\sqrt{\frac{nm}{n+m}}(\hat{C}^{\pi}_{1,n}-\hat{C}^{\pi}_{2,m})$. The computation based on the latter permutation statistic provides a valid approximation of the limit of our test statistic whereas the former permutation statistic can only be used to approximate $\mathbb{\tilde{T}}$.

\textit{\textbf{Remark 3.1.}} An important distinction has to be made between the copula
associated with the mixture distribution and the mixture copula.
In Section 2, we argued that the copula
associated with the mixture distribution $\bar{H}=\lambda H_1+(1-\lambda)H_2$ is not necessarily equal to $C_{1}$ (or $C_{2}$%
) even when $C_{1}=C_{2}$. On the other hand, the mixture copula $\bar{C}=\lambda C_1+(1-\lambda)C_2$ is
always equal to $C_{1}$ (or $C_{2}$) whenever $C_{1}=C_{2}$ for any $%
\lambda \in \lbrack 0,1]$. The latter property suggests that the
permutations of $Z$ in (8) properly reflect the group
invariance conditions implied by copula equality.

\textit{\textbf{Remark 3.2.}} Under the condition that $F_{i}=G_{i}$ with $F_{i}^{\prime}>0$ for all $i=1,...,d$, the permutation test based on the permutations of $W$ in (2) is also valid with the test statistic $T_{n,m}^{(p)}$. This is due to the invariance
property of copula, which says that for any strictly increasing transformations $\beta _{i}$ for $i=1,...,d$, the copula of $(\beta _{1}(X_{1}),...,\beta _{d}(X_{d}))$ is equivalent to the the copula of $(X_{1},...,X_{d})$.
In light of the discussion in Section
2, note that $H_{1}=H_{2}$ if and only if (1) $F_{i}=G_{i}$ for all $%
i=1,...,d$ and (2) $C_{1}=C_{2}$. Therefore, under the condition that $%
F_{i}=G_{i}$ for all $i=1,...,d$, the two sets $\Theta_{0} $ and $\Theta _{00}$
are equal.

\textit{\textbf{Remark 3.3.}}
In related literature, Canay et al. (2017) investigate randomization tests under
approximate group invariance conditions, satisfied when there is a map $S_{N}$ from a sample space $%
\mathcal{W}_{N}$ to $\mathcal{S}$ such that (i) $S_{N}(W)$ for $W\in \mathcal{W}_{N}$ converges in
distribution to $S$ $\in \mathcal{S}$ as $N\rightarrow \infty $, and (ii)
$gS\buildrel d\over= S$ for all $g$ in some finite group of transformations $\mathbf{G}$. Our setting is somewhat different.
For us, $S_{N}$ can be defined by the map which transforms $W$ into $%
\hat{Z}$, where $\hat{Z}$ is an approximation to $Z$. Then an approximate group invariance holds by the distributional equality $g(Z)\buildrel d\over=Z$ for any $N$ and $g\in \mathbf{G}_{N}$. Since $\mathcal{S}$ and $\mathbf{G}$ in this context depend on $N$, our approximate group invariance condition cannot be
reformulated in the framework of Canay et al. (2017) and should be handled
in a different manner.

\textit{\textbf{Remark 3.4.}} Beare and Seo (2017) also explore an asymptotic group invariance condition to develop a quasi-randomization test of copula symmetry. While $g\in \mathbf{%
G}_{N}$ in our setting is specified by a permutation $\pi $ which is
uniformly distributed over $\mathbf{G}_{N}$, $g\in \mathbf{G}_{n}$ in Beare
and Seo (2017) is defined to be a transformation from $([0,1]^{2})^{n}$ onto
itself defined by $g((x_{1},y_{1}),...,(x_{n},y_{n}))=(\pi ^{\tau
_{1}}(x_{1},y_{1}),...,\pi ^{\tau _{n}}(x_{n},y_{n}))$ where $(\pi
^{0}(x,y),\pi ^{1}(x,y))=((x,y),(y,x))$ or $(\pi ^{0}(x,y),\pi
^{1}(x,y))=((x,y),(1-x,1-y))$ with $\tau _{i}$ being $n$ i.i.d. draws from the
Bernoulli distribution. Unlike our framework, the group invariance conditions in Beare and Seo (2017) hold between paired normalized
observations (with $n=m$) which are completely dependent with each other. Hence, our proofs are considerably different from theirs though some results on the conditional convergence in Beare and Seo (2017) are also useful to us.

\textit{\textbf{Remark 3.5.}} It should not be overlooked that the assumptions on $\lambda _{n,m}$ have been strengthened to obtain desired results. For Lemma 3.1, we only require $%
\lambda _{n,m}$ to converge to $\lambda $ as $\min (n,m)\rightarrow \infty $. For Lemma 3.2,
Lemma 3.3 and Theorem 3.1, $\lambda _{n,m}$ should satisfy a certain
convergence rate $\lambda _{n,m}-\lambda =O((n+m)^{-1/2})$ for applications of contiguity argument and
coupling construction in Chung and Romano (2013). The assumption in Lemma 3.4 and Theorem
3.2 is stronger, requiring that $\lambda _{n,m}-\lambda $ decays
faster than $(\min (n,m))^{-1/2}$.

\section{Simulations}
Here we report some Monte Carlo simulation results to show the
finite sample performance of our proposed test. We particularly examine the two cases, $p=2$ and $p=\infty $ for the choice of $p$, which lead to the Cram\'er von-Mises statistic and Kolmogorov-Smirnov statistic respectively. We distinguish them by using the different notations $T_{n,m}^{(2)}$ and $T_{n,m}^{(\infty )}$. $T_{n,m}^{(2)}$ can be computed using the formula
\begin{eqnarray*}
(T_{n,m}^{(2)})^{2} &=&\frac{nm}{n+m}\left\{ \frac{1}{n^{2}}%
\sum_{i=1}^{n}\sum_{j=1}^{n}\prod\limits_{q=1}^{d}\min ( 1-\hat{U}_{q,n}^{i},1-\hat{U}_{q,n}^{j}) \right.  \\
&&\left. +\frac{1}{m^{2}}\sum_{i=1}^{m}\sum_{j=1}^{m}\prod\limits_{q=1}^{d}%
\min ( 1-\hat{V}_{q,m}^{i},1-\hat{V}_{q,m}^{j}) -\frac{2}{nm}%
\sum_{i=1}^{n}\sum_{j=1}^{m}\prod\limits_{q=1}^{d}\min ( 1-\hat{U}%
_{q,n}^{i},1-\hat{V}_{q,m}^{j}) \right\},  \\
\end{eqnarray*}
while $T_{n,m}^{(\infty )}$ can be calculated as
\begin{eqnarray*}
T_{n,m}^{(\infty )} &=&\sqrt{\frac{nm}{n+m}}\max_{S_{1}\cup S_{2}}\left\vert \hat{C}_{1,n}\left(%
\frac{i_{1}}{n},...,\frac{i_{d}}{n}\right)-\hat{C}_{2,m}\left(\frac{j_{1}}{m},...,\frac{%
j_{d}}{m}\right)\right\vert \text{ with} \\
S_{1} &=&\left\{ (i_{1},...,i_{d},j_{1},...,j_{d})|j_{q}=\left[\frac{m}{n}%
i_{q}\right],i_{q}=1,2,...n\text{ for }q=1,...,d\right\} \text{ and} \\
S_{2} &=&\left\{ (i_{1},...,i_{d},j_{1},...,j_{d})|i_{q}=\left[\frac{n}{m}%
j_{q}\right],j_{q}=1,2,...m\text{ for }q=1,...,d\right\}.
\end{eqnarray*}%
We found that the test proposed in Theorem 3.1 generally leads to similar results with slightly less power than the test proposed in Theorem 3.2. Hence, we only display the rejection frequencies of the test proposed in Theorem 3.2 in this section. Throughout the simulation, we employed $1000$ replications, and in each replication we conducted $1000$
random permutations to obtain the critical values. Random
sampling of the permutation does not change the asymptotic properties
established in Section 3 as the number of random permutations approaches infinity.

\begin{table}[t!]
	\footnotesize
\begin{center}
\begin{tabular}{c|c|l|ccccccc}
\hline\hline
$\alpha$ &  Test &$(n,m)$ & Gaussian & Student-t & Clayton & Frank & Gumbel & Sym-JC & Plackett\\
\hline\hline
\multirow{9}{*}{$0.05$} & \multirow{3}{*}{$R_{n,m}^{(2)}$} &$(5,10)$ &0.268 &0.243 &0.244 &0.264 &0.261 &0.249 &0.242\\
\multirow{9}{*}{}&\multirow{3}{*}{} &$(10,5)$ &0.249 &0.243 &0.248 &0.263 &0.286 &0.241 &0.260\\
\multirow{9}{*}{}&\multirow{3}{*}{} &$(50,50)$ &0.045 &0.037 &0.048 &0.037 &0.035 &0.043 &0.044\\
\cline{2-10} 

\multirow{9}{*}{}&\multirow{3}{*}{$T_{n,m}^{(2)}$} &$(5,10)$ &0.048 &0.050 &0.045 &0.048 &0.044 &0.046 &0.045\\
\multirow{9}{*}{}&\multirow{3}{*}{} &$(10,5)$ &0.049 &0.051 &0.047 &0.045 &0.052 &0.050 &0.040\\
\multirow{9}{*}{}&\multirow{3}{*}{} &$(50,50)$ &0.046 &0.046 &0.042 &0.045 &0.043 &0.047 &0.040\\
\cline{2-10} 

\multirow{9}{*}{}&\multirow{3}{*}{$T_{n,m}^{(\infty )}$} &$(5,10)$ &0.047 &0.045 &0.045 &0.044 &0.040 &0.047 &0.054\\
\multirow{9}{*}{}&\multirow{3}{*}{} &$(10,5)$ &0.051 &0.048 &0.040 &0.044 &0.050 &0.049 &0.046\\
\multirow{9}{*}{}&\multirow{3}{*}{} &$(50,50)$ &0.043 &0.045 &0.045 &0.046 &0.041 &0.048 &0.042\\

\cline{1-10} 

\multirow{9}{*}{$0.1$} & \multirow{3}{*}{$R_{n,m}^{(2)}$} &$(5,10)$ &0.502 &0.496 &0.467 &0.508 &0.506 &0.493 &0.503\\
\multirow{9}{*}{}&\multirow{3}{*}{} &$(10,5)$ &0.466 &0.474 &0.490 &0.497 &0.527 &0.469 &0.485\\
\multirow{9}{*}{}&\multirow{3}{*}{} &$(50,50)$ &0.083 &0.075 &0.092 &0.081 &0.070 &0.088 &0.084\\
\cline{2-10} 

\multirow{9}{*}{} & \multirow{3}{*}{$T_{n,m}^{(2)}$} &$(5,10)$ &0.086 &0.087 &0.095 &0.089 &0.076 &0.081 &0.098\\
\multirow{9}{*}{}&\multirow{3}{*}{} &$(10,5)$ &0.099 &0.090 &0.102 &0.083 &0.092 &0.099 &0.090\\
\multirow{9}{*}{}&\multirow{3}{*}{} &$(50,50)$ &0.084 &0.079 &0.086 &0.079 &0.075 &0.083 &0.078\\
\cline{2-10} 

\multirow{9}{*}{} & \multirow{3}{*}{$T_{n,m}^{(\infty )}$} &$(5,10)$ &0.102 &0.089 &0.084 &0.078 &0.079 &0.086 &0.100\\
\multirow{9}{*}{}&\multirow{3}{*}{} &$(10,5)$ &0.094 &0.082 &0.089 &0.089 &0.098 &0.085 &0.088\\
\multirow{9}{*}{}&\multirow{3}{*}{} &$(50,50)$ &0.085 &0.084 &0.082 &0.078 &0.077 &0.087 &0.090\\
\cline{1-10} 

\multirow{9}{*}{$0.2$} & \multirow{3}{*}{$R_{n,m}^{(2)}$} &$(5,10)$ &0.829 &0.832 &0.800 &0.830 &0.842 &0.835 &0.828\\
\multirow{9}{*}{}&\multirow{3}{*}{} &$(10,5)$ &0.815 &0.805 &0.789 &0.835 &0.872 &0.837 &0.814\\
\multirow{9}{*}{}&\multirow{3}{*}{} &$(50,50)$ &0.168 &0.173 &0.179 &0.161 &0.148 &0.164 &0.175\\
\cline{2-10} 

\multirow{9}{*}{}&\multirow{3}{*}{$T_{n,m}^{(2)}$} &$(5,10)$ &0.179 &0.178 &0.188 &0.202 &0.181 &0.195 &0.202\\
\multirow{9}{*}{}&\multirow{3}{*}{} &$(10,5)$ &0.200 &0.189 &0.203 &0.179 &0.190 &0.203 &0.199\\
\multirow{9}{*}{}&\multirow{3}{*}{} &$(50,50)$ &0.172 &0.180 &0.188 &0.173 &0.175 &0.177 &0.173\\
\cline{2-10} 

\multirow{9}{*}{}&\multirow{3}{*}{$T_{n,m}^{(\infty )}$} &$(5,10)$ &0.188 &0.179 &0.171 &0.173 &0.182 &0.169 &0.181\\
\multirow{9}{*}{}&\multirow{3}{*}{} &$(10,5)$ &0.199 &0.177 &0.188 &0.189 &0.173 &0.184 &0.174\\
\multirow{9}{*}{}&\multirow{3}{*}{} &$(50,50)$ &0.185 &0.186 &0.179 &0.176 &0.164 &0.174 &0.176\\
\cline{1-10} 
\hline\hline
\end{tabular}%
\end{center}
\caption{The table shows the null rejection frequencies of our tests, $T_{n,m}^{(2)}$ and $T_{n,m}^{(\infty)}$, and the bootstrap test, $R_{n,m}^{(2)}$. The marginal distributions are taken to be uniform. Nominal level is set to be $\alpha=(0.05, 0.1, 0.2)$.}
\end{table}

We first generate two groups of i.i.d. samples from the same copula to examine the null rejection rates of our tests. For the choice of copula, we employ bivariate Gaussian, Student-t, Clayton, Frank, Gumbel, symmetrized Joe-Clayton and Plackett copulas. The parameter of each copula is fixed so as to provide the same level of dependence, which yields the correlation coefficient $0.5$ in the Gaussian case. The dependence structures of these copulas are displayed in Figure 1 of Patton (2006). 
As we observe in Table 4.1, the computed rejection frequencies with $T_{n,m}^{(2)}$ and $T_{n,m}^{(\infty )}$ are very close to the nominal level with the sample size of $(n,m)=(5,10)$ or $(10,5)$, while the bootstrap test by R\'{e}millard and Scaillet (2009) tends to overreject in small samples. Although our test procedure has been justified asymptotically, the table shows that it has excellent size control in finite samples. On the other hand, both permutation and bootstrap tests control size well when the sample size is over $50$ in each group.

\begin{table}[!htbp]
	\footnotesize
\begin{center}
\begin{tabular}{c|c|l|ccccccc}
\hline\hline
Model & Test & $(n,m)$ & $\tau_{2}=0.3$ & $\tau_{2}=0.4$ & $\tau_{2}=0.5$ & $\tau_{2}=0.6$ & $\tau_{2}=0.7$ & $\tau_{2}=0.8$ & $\tau_{2}=0.9$\\
\hline\hline
\multirow{6}{*}{Gaussian}& \multirow{2}{*}{$R_{n,m}^{(2)}$}&(10,50) &  0.054   &  0.068   & 0.131  &  0.220  &  0.327 &  0.442  &  0.531\\
\multirow{6}{*}{}&\multirow{2}{*}{}&(50,100) &0.122 &0.362 &0.707 &0.946 &0.997 &1.000  &  1.000\\
\cline{2-10} 
\multirow{6}{*}{}&\multirow{2}{*}{$T_{n,m}^{(2)}$} &(10,50) &   0.111     &0.179    &0.324    &0.516    &0.706    &0.873  & 0.973\\
\multirow{6}{*}{}&\multirow{2}{*}{}&(50,100) &0.152  &0.448 &0.815 &0.977 &1.000 &1.000  & 1.000\\
\cline{2-10} 
\multirow{6}{*}{}&\multirow{2}{*}{$T_{n,m}^{(\infty)}$} &(10,50) &  0.093 &  0.170   & 0.270 &   0.432  &  0.601  &  0.821  &  0.932\\
\multirow{6}{*}{}&\multirow{2}{*}{}&(50,100)  &0.134 &0.362 &0.691 &0.910 &0.993 &1.000  & 1.000\\
\hline 

\multirow{6}{*}{Student-t}& \multirow{2}{*}{$R_{n,m}^{(2)}$}&(10,50) & 0.055   &    0.075 &   0.117  &  0.178  &  0.260  &  0.353  &  0.463\\
\multirow{6}{*}{}&\multirow{2}{*}{}&(50,100) &0.099  &0.320 &0.655 &0.912 &0.989 &1.000  &1.000\\
\cline{2-10} 
\multirow{6}{*}{}&\multirow{2}{*}{$T_{n,m}^{(2)}$} &(10,50)     &0.102      &0.174    &0.317    &0.467    &0.634    &0.814  & 0.957\\
\multirow{6}{*}{}&\multirow{2}{*}{}&(50,100) &0.143  &0.408 &0.745 &0.950 &0.996 &1.000  & 1.000\\
\cline{2-10} 
\multirow{6}{*}{}&\multirow{2}{*}{$T_{n,m}^{(\infty)}$} &(10,50) & 0.106   &  0.153  &  0.266  &  0.380  &  0.571  &  0.758  & 0.916\\
\multirow{6}{*}{}&\multirow{4}{*}{}&(50,100) &0.147 &0.354 &0.645 &0.891 &0.988 &1.000  & 1.000\\
\hline 

\multirow{6}{*}{Clayton}& \multirow{2}{*}{$R_{n,m}^{(2)}$}&(10,50)& 0.056     & 0.074    &0.118  &  0.182    &0.284   & 0.400  & 0.476\\
\multirow{6}{*}{}&\multirow{2}{*}{}&(50,100) &0.111 &0.370 &0.736 &0.949 &0.998 &1.000  & 1.000\\
\cline{2-10} 
\multirow{6}{*}{}&\multirow{2}{*}{$T_{n,m}^{(2)}$} &(10,50) &0.110 &0.178 &0.296 &0.491 &0.684 &0.847  & 0.956\\
\multirow{6}{*}{}&\multirow{2}{*}{}&(50,100) &0.174 &0.426 &0.799 &0.975 &1.000 &1.000  & 1.000\\
\cline{2-10} 
\multirow{6}{*}{}&\multirow{2}{*}{$T_{n,m}^{(\infty)}$} &(10,50) & 0.099   &  0.141   & 0.244   & 0.385 &   0.571  &  0.774  & 0.911\\
\multirow{6}{*}{}&\multirow{4}{*}{}&(50,100) &0.145  &0.326 &0.660 &0.901 &0.995 &0.999  & 1.000\\
\hline 

\multirow{6}{*}{Frank}& \multirow{2}{*}{$R_{n,m}^{(2)}$}&(10,50) & 0.055      & 0.074 &   0.126  & 0.215   & 0.308 &   0.424  &  0.542\\
\multirow{6}{*}{}&\multirow{2}{*}{}&(50,100) &0.124  &0.392 &0.764 &0.974 &1.000 &1.000  & 1.000\\
\cline{2-10} 
\multirow{6}{*}{}&\multirow{2}{*}{$T_{n,m}^{(2)}$} &(10,50) &0.101  &0.214 &0.353 &0.541 &0.751 &0.886  & 0.967\\
\multirow{6}{*}{}&\multirow{2}{*}{}&(50,100) &0.160 &0.453 &0.839 &0.980 &1.000 &1.000  & 1.000\\
\cline{2-10} 
\multirow{6}{*}{}&\multirow{2}{*}{$T_{n,m}^{(\infty)}$} &(10,50) &  0.110   &0.178   & 0.309&    0.425   & 0.670 &   0.836  & 0.935\\
\multirow{6}{*}{}&\multirow{4}{*}{}&(50,100)  &0.130 &0.378 &0.731 &0.955 &0.996 &1.000  & 1.000\\
\hline 

\multirow{6}{*}{Gumbel}& \multirow{2}{*}{$R_{n,m}^{(2)}$}&(10,50) &0.058   &   0.085  &   0.123  &   0.198  &   0.287   & 0.414  &  0.531\\
\multirow{6}{*}{}&\multirow{2}{*}{}&(50,100) &0.113 &0.345 &0.712 &0.943 &1.000 &1.000  & 1.000\\
\cline{2-10} 
\multirow{6}{*}{}&\multirow{2}{*}{$T_{n,m}^{(2)}$} &(10,50) &0.097  &0.165 &0.305 &0.482 &0.661 &0.841  & 0.959\\
\multirow{6}{*}{}&\multirow{2}{*}{}&(50,100) &0.142 &0.404 &0.771 &0.966 &0.999 &1.000  & 1.000\\
\cline{2-10} 
\multirow{6}{*}{}&\multirow{2}{*}{$T_{n,m}^{(\infty)}$} &(10,50) &   0.104  &   0.157   & 0.255 &   0.405  &  0.579   & 0.789  & 0.916\\
\multirow{6}{*}{}&\multirow{4}{*}{}&(50,100)  &0.125 &0.335 &0.672 &0.905 &0.990 &1.000  & 1.000\\
\hline 

\multirow{6}{*}{Sym-JC}&\multirow{2}{*}{$R_{n,m}^{(2)}$}&(10,50)& 0.054    & 0.071  &  0.106  &   0.154  & 0.231  &  0.323  & 0.410\\
\multirow{6}{*}{}&\multirow{2}{*}{}&(50,100) &0.105 &0.391 &0.717 &0.943 &0.997 &1.000  & 1.000\\
\cline{2-10} 
\multirow{6}{*}{}&\multirow{2}{*}{$T_{n,m}^{(2)}$}&(10,50) &0.094 &0.195 &0.311 &0.477 &0.660 &0.851  & 0.935\\
\multirow{6}{*}{}&\multirow{2}{*}{}&(50,100) &0.146 &0.469 &0.794 &0.964 &1.000 &1.000  & 1.000\\
\cline{2-10} 
\multirow{6}{*}{}&\multirow{2}{*}{$T_{n,m}^{(\infty)}$} &(10,50) &   0.103  &  0.200  &  0.286  &  0.446  &  0.652  &  0.823  & 0.864\\
\multirow{6}{*}{}&\multirow{4}{*}{}&(50,100)  &0.145 &0.423 &0.721 &0.931 &0.999 &1.000  & 1.000\\
\hline 

\multirow{6}{*}{Plackett}&\multirow{2}{*}{$R_{n,m}^{(2)}$}&(10,50)  &0.065    &  0.086   & 0141 &   0.217  &  0.324  &  0.417  &  0.524\\
\multirow{6}{*}{}&\multirow{2}{*}{}&(50,100)& 0.116  &	0.396	&0.744	&0.944	&0.995	&1.000  & 1.000\\
\cline{2-10} 
\multirow{6}{*}{}&\multirow{2}{*}{$T_{n,m}^{(2)}$}&(10,50) &0.086  &0.189 &0.319 &0.458 &0.648 &0.811  & 0.945\\
\multirow{6}{*}{}&\multirow{2}{*}{}&(50,100) &0.170		&0.455	&0.798	&0.972	&1.000	&1.000  & 1.000\\
\cline{2-10} 
\multirow{6}{*}{}&\multirow{2}{*}{$T_{n,m}^{(\infty)}$} &(10,50) &  0.087  & 0.162  &  0.278&    0.402   & 0.603   & 0.760  & 0.916\\
\multirow{6}{*}{}&\multirow{4}{*}{}&(50,100) &0.143	&	0.398	&0.721&	0.918&	0.997	&1.000  & 1.000\\
\hline\hline 
\end{tabular}%
\end{center}
\caption{This table shows the power of our tests with $T_{n,m}^{(2)}$ and $T_{n,m}^{(\infty)}$, and the bootstrap test, $R_{n,m}^{(2)}$. The marginal distributions are taken to be uniform. Nominal level is set to be $\alpha=0.05$.}
\end{table}

Next, we provide Table 4.2 to illustrate the power of the tests for copula equality. We follow the simulation design of R\'{e}millard and Scaillet (2009) and select the copulas $C_{1}$ and $C_{2}$ from the same copula family, but with possibly different copula parameters. More specifically, we fix the Kendall's $\tau $ of $C_{1}$ to be $\tau_1=0.2$, while we let the Kendall's $\tau $ of $C_{2}$ vary over the range $\tau_{2}\in \{0.3, 0.4,0.5,0.6,0.7,0.8, 0.9\}$. Note that the results for our test with the statistic $T_{n,m}^{(2)}$ can be directly comparable to the results for the boostrap test in R\'{e}millard and Scaillet (2009) because the two tests use the same test statistic but different critical values. Since the null rejection rates of the bootstrap test are found to be between $0.12$ and $0.15$ with the sample size of $(n,m)=(10,50)$, we report size-adjusted power of the bootstrap test for this sample size. When the sample size is small, we find that the power of permutation test dominates that of bootstrap test, while the rejection frequencies are similar when the sample size is large. This phenomenon can be understood in the same context as Romano (1989).

\begin{table}[!htbp]
	\footnotesize
\begin{center}
\begin{tabular}{c|c|l|ccccccc}
\hline\hline
Model & Test & $(n,m)$ & $\tau_{2}=0.3$& $\tau_{2}=0.4$ & $\tau_{2}=0.5$ & $\tau_{2}=0.6$ & $\tau_{2}=0.7$ & $\tau_{2}=0.8$ & $\tau_{2}=0.9$ \\
\hline\hline
\multirow{6}{*}{Gaussian}&\multirow{2}{*}{$R_{n,m}^{(2)}$}&(10,50) &   0.053  &   0.074&     0.130 &    0.216  &   0.329 &    0.451  & 0.540\\
\multirow{6}{*}{}&\multirow{2}{*}{}&(50,100) &0.115  &0.411 &0.744 &0.947 &0.996 &1.000  & 1.000\\
\cline{2-10} 
\multirow{6}{*}{}&\multirow{2}{*}{$T_{n,m}^{(2)}$}	&(10,50)   &  0.111 &  0.206   & 0.334  &  0.502 &   0.710  &  0.880  & 0.973\\
\multirow{6}{*}{}&\multirow{4}{*}{}&(50,100) &0.152 &0.435 &0.790 &0.975 &0.999 &1.000  & 1.000\\
\cline{2-10} 
\multirow{6}{*}{}&\multirow{2}{*}{$T_{n,m}^{(\infty)}$} &(10,50) & 0.093    & 0.168  &  0.270   & 0.412   & 0.619  &  0.812  & 0.932\\
\multirow{6}{*}{}&\multirow{4}{*}{}&(50,100) &0.134 &0.329 &0.646 &0.908 &0.988 &1.000  & 1.000\\
\hline 

\multirow{6}{*}{Student-t}&\multirow{2}{*}{$R_{n,m}^{(2)}$}&(10,50) &    0.064 &  0.081  &  0.124   & 0.196  &  0.284 &   0.391  &  0.495\\
\multirow{6}{*}{}&\multirow{2}{*}{}&(50,100) &0.100 &0.341 &0.609 &0.893 &0.990 &0.999  & 1.000\\
\cline{2-10} 
\multirow{6}{*}{}&\multirow{2}{*}{$T_{n,m}^{(2)}$}	&(10,50) &  0.089 &  0.182 &   0.290   & 0.480  &  0.675   & 0.847  & 0.941\\
\multirow{6}{*}{}&\multirow{2}{*}{}&(50,100) &0.143&0.408 &0.745 &0.950 &0.996 &1.000  & 1.000\\
\cline{2-10} 
\multirow{6}{*}{}&\multirow{2}{*}{$T_{n,m}^{(\infty)}$} &(10,50)   &   0.083  &  0.162  &  0.248 &   0.402 &   0.573 &   0.769  & 0.904\\
\multirow{6}{*}{}&\multirow{4}{*}{}&(50,100)&0.147 &0.354 &0.645 &0.891 &0.988 &1.000  & 1.000\\
\hline 
	
\multirow{6}{*}{Clayton}&\multirow{2}{*}{$R_{n,m}^{(2)}$}&(10,50) &    0.056  &  0.077  &  0.129  &  0.203  &  0.292  &  0.397  & 0.485\\
\multirow{6}{*}{}&\multirow{2}{*}{}&(50,100) &0.130  &0.373 &0.716 &0.948 &0.998 &1.000  & 1.000\\
\cline{2-10} 
\multirow{6}{*}{}&\multirow{2}{*}{$T_{n,m}^{(2)}$}	&(10,50)  &  0.103  &  0.202  &  0.341 &   0.504  &  0.694  &  0.874  & 0.951\\
\multirow{6}{*}{}&\multirow{2}{*}{}&(50,100) &0.147 &0.426 &0.799 &0.975 &1.000 &1.000  & 1.000\\
\cline{2-10} 
\multirow{6}{*}{}&\multirow{2}{*}{$T_{n,m}^{(\infty)}$} &(10,50)  &  0.103  &  0.163 &   0.267 &   0.421   & 0.588  &  0.797  & 0.907\\
\multirow{6}{*}{}&\multirow{4}{*}{}&(50,100) &0.142 &0.351 &0.631 &0.879 &0.983 &0.999 & 1.000\\
\hline 

\multirow{6}{*}{Frank}&\multirow{2}{*}{$R_{n,m}^{(2)}$}&(10,50) &0.053  &   0.066 &   0.118  &  0.196  &  0.290  &  0.377  & 0.466\\
\multirow{6}{*}{}&\multirow{2}{*}{}&(50,100) &0.126  &0.389 &0.795 &0.976 &1.000 &1.000  & 1.000\\
\cline{2-10} 
\multirow{6}{*}{}&\multirow{2}{*}{$T_{n,m}^{(2)}$}	&(10,50)  &  0.101 &  0.201  &  0.335  &  0.538  &  0.737  &  0.875  & 0.967\\
\multirow{6}{*}{}&\multirow{2}{*}{}&(50,100) &0.160&0.485 &0.849 &0.990 &1.000 &1.000  &1.000\\
\cline{2-10} 
\multirow{6}{*}{}&\multirow{2}{*}{$T_{n,m}^{(\infty)}$} &(10,50)&0.110  &   0.175  &  0.274  &  0.446  &  0.647 &   0.815  & 0.935\\
\multirow{6}{*}{}&\multirow{4}{*}{}&(50,100)&0.130  &0.392 &0.750 &0.955 &1.000 &1.000  & 1.000\\
\hline 

\multirow{6}{*}{Gumbel}&\multirow{2}{*}{$R_{n,m}^{(2)}$}&(10,50) &0.054 &   0.063 &   0.107 &   0.177  &  0.270 &   0.380  & 0.465\\
\multirow{6}{*}{}&\multirow{2}{*}{}&(50,100) &0.108  &0.357 &0.715 &0.931 &0.996 &1.000  & 1.000\\
\cline{2-10} 
\multirow{6}{*}{}&\multirow{2}{*}{$T_{n,m}^{(2)}$}	&(10,50)   &  0.097   &  0.189  &  0.315   & 0.479   & 0.688  &  0.868  & 0.959\\
\multirow{6}{*}{}&\multirow{2}{*}{}&(50,100) &0.142 &0.437 &0.790 &0.975 &0.999 &1.000  & 1.000\\
\cline{2-10} 
\multirow{6}{*}{}&\multirow{2}{*}{$T_{n,m}^{(\infty)}$} &(10,50) & 0.104   & 0.157   & 0.264 &   0.408   & 0.585  &  0.798  & 0.916\\
\multirow{6}{*}{}&\multirow{4}{*}{}&(50,100) &0.125 &0.365 &0.670 &0.919 &0.993 &1.000  & 1.000\\
\hline 

\multirow{6}{*}{Sym-JC}&\multirow{2}{*}{$R_{n,m}^{(2)}$}&(10,50) &  0.055  &    0.066  &   0.090  &   0.131  &   0.203 &    0.278  & 0.350\\
\multirow{6}{*}{}&\multirow{2}{*}{}&(50,100) &0.113 &0.384 &0.733 &0.945 &0.998 &1.000  & 1.000\\
\cline{2-10} 
\multirow{6}{*}{}&\multirow{2}{*}{$T_{n,m}^{(2)}$}	&(10,50)  &  0.103    &  0.175  &  0.288  &  0.442  &  0.637  &  0.798  & 0.930\\
\multirow{6}{*}{}&\multirow{2}{*}{}&(50,100) &0.133  &0.428 &0.792 &0.967 &0.999 &1.000  & 1.000\\
\cline{2-10} 
\multirow{6}{*}{}&\multirow{2}{*}{$T_{n,m}^{(\infty)}$} &(10,50) &0.101  &  0.170  &  0.258 &   0.397  &  0.602 &   0.797  & 0.865\\
\multirow{6}{*}{}&\multirow{4}{*}{}&(50,100) &0.131 &0.395 &0.728 &0.933 &0.998 &1.000  & 1.000\\
\hline 

\multirow{6}{*}{Plackett}&\multirow{2}{*}{$R_{n,m}^{(2)}$}&(10,50) &0.068   &  0.100  &  0.163  &  0.233  &  0.343  &  0.448  & 0.557\\
\multirow{6}{*}{}&\multirow{2}{*}{}&(50,100) &0.122	&0.409	&0.738	&0.944	&0.997&	1.000  & 1.000\\
\cline{2-10} 
\multirow{6}{*}{}&\multirow{2}{*}{$T_{n,m}^{(2)}$}	&(10,50)  &  0.093 &   0.185  &  0.320  &  0.458   & 0.651  &  0.816  & 0.939\\
\multirow{6}{*}{}&\multirow{2}{*}{}&(50,100)& 0.160&	0.458	&0.803	&0.966&	1.000	&1.000  & 1.000\\
\cline{2-10} 
\multirow{6}{*}{}&\multirow{2}{*}{$T_{n,m}^{(\infty)}$} &(10,50)& 0.086    &  0.161   & 0.264   & 0.397 &   0.571   & 0.751  & 0.910\\
\multirow{6}{*}{}&\multirow{4}{*}{}&(50,100) &0.152	&0.417	&0.730	&0.910&	0.998	&1.000  & 1.000\\
\hline\hline 
\end{tabular}%
\end{center}
\caption{This table shows the power of our tests with $T_{n,m}^{(2)}$ and $T_{n,m}^{(\infty)}$, and the bootstrap test, $R_{n,m}^{(2)}$. The marginal distributions are taken to be normal. Nominal level is set to be $\alpha=0.05$.}
\end{table}

Lastly, we conduct a simulation study using different marginal distributions from the uniform distribution. The two margins of $H_{1}$, namely $F_{1}$ and $F_{2}$, are taken from $N(0,1)$ while those of $H_{2}$, $G_{1}$ and $G_{2}$, are taken from $N(5,1)$. In Table 4.3, we find that changing the margins leads to no appreciable difference in rejection rates. This reflects the fact that the asymptotic property of our tests is not affected by such changes in the margins, as it should be. On the other hand, permutation tests based on the permutations of the original samples (i.e., permutations of the rows of $W$) do not yield the correct size, as we discussed in Section 2. Under the same simulation setup, we found that these tests over-reject the null hypothesis with the rejection rates reaching almost $0.3$, with the sample size  $(n,m)=(100,100)$. Increasing the sample size does not improve the size of the tests. We also performed a simulation study based on the different setting of $F_{1}=G_{1} \sim N(0,1)$ and $F_{2}=G_{2}\sim N(5,1)$. Under this framework, both of the randomization tests by permuting the rows of $W$ and the rows of $\hat{Z}$ are valid by Remark 3.2. We do not report these additional results separately because the rejection rates are very similar to those in Table 4.2 (or Table 4.3).



\section{Empirical applications}
\subsection{Dependence of income and consumption}
Household consumption decisions are of central concern in economics. In the short run, fluctuations in consumption induce business cycles while in the long run, consumption behavior serves as a primary determinant of economic growth. In the classical model of Keynes (1936), consumption is represented as a function of income. Although there can be other determinants of consumption, substantial empirical evidence shows that disposable income plays the most important role in explaining consumer behavior. See Hall (1978), Flavin (1981), Hall and Mishkin (1982), Campbell and Deaton (1989), Shapiro and Slemrod (1995), Shea (1995), Parker (1999), Souleles (1999), Johnson et al.\ (2006), Parker et al.\ (2013) and Kaplan and Violante (2014) for research along these lines.\footnote{There has been a controversial debate on the relationship between income and consumption. According to the Permanent Income Hypothesis (Friedman, 1957), individual income consists of transitory income and permanent income, and consumption is determined by the permanent income component rather than the transitory income component. In a similar context, the Life Cycle Hypothesis (Ando and Modigliani, 1963) assumes that the utility of an individual consumer depends on his own total consumption in current and future periods, and utility maximization under intertemporal budget constraints yields the solution of current consumption expressed in terms of the total resources over the life time and the rate of capital return. From the perspective of the Permanent Income Hypothesis or Life Cycle Hypothesis, current consumption should not be affected by a change in transitory income or anticipated income. See Sargent (1978), Browning and Collado (2001) and Hsieh (2003) for related empirical evidence.} In this section, we estimate copulas to study how consumption relates to disposable income with a focus on their dependence structure. By applying the proposed test of copula equality, we examine whether the dependence structure is identical in several different circumstances.

Since our goal is to study the structure of dependence, we explore micro-level household data of income and consumption in several countries at different stages of economic development. For the cross-country analysis, we investigate household annual income and annual consumption in 2010 collected from the U.S., Mexico and South Africa.
In each survey, we compute household income by summing over the annual income from work, transfers, rental, and other miscellaneous income (including public assistance), after deducting the taxes. For the household consumption, we use the total sum of the expenses that households made on the food, clothing, housing, health care, education, transportation, trip, furniture and equipment, entertainment and other miscellaneous expenditure for a year. The sample sizes of the data for the U.S., Mexico, and South Africa are 16,803, 27,614 and 25,243, respectively.\footnote{Data sources are \href{https://www.bls.gov/cex/pumd_data.htm}{the U.S. Consumer Expenditure Survey} (https://www.bls.gov/cex/pumddata.htm), \href{http://www.beta.inegi.org.mx/proyectos/enchogares/regulares/enigh/tradicional/2010/default.html}{Mexico Household Income and Expenditure Survey} (https://www.beta.inegi.org.mx/proyectos/enchogares
/regulares/enigh/tradicional/2010/default.html) and \href{https://www.datafirst.uct.ac.za/dataportal/index.php/catalog/316}{Income and Expenditure Survey in South Africa} (https://www.datafirst.uct.ac.za/dataportal/index.php/catalog/316).}

To provide an overview of the dependence structure between income and consumption, we apply the probability integral transforms to the data of the three countries and display the scatterplots in Figure 5.1. We observe that the dependence of income and consumption is stronger in South Africa and Mexico than in the U.S. To be precise, the correlation coefficients of income and consumption in the U.S., Mexico, and South Africa, are 0.54, 0.74 and 0.76, while the Kendall's $\tau$ are $0.62$, $0.78$ and $0.76$ respectively. In general, the ratio of consumption to income is relatively higher in developing countries, which leads to stronger positive dependence in the relation of income and consumption. In addition, consumers face lack of credit and insurance in developing countries, and this results in less consumption smoothing over the life time. As a consequence, the dependence of current income and current consumption may tend to be stronger in developing countries (See Jappelli and Pagano; 1989, Campbell and Mankiw; 1991, Rosenzweig and Wolpin; 1993, Zimmerman and Carter; 2003, Gin\'{e} and Yang; 2009 and Karlan et al.\ (2014)). 

\begin{figure}[t!]
  \hspace{-3cm}
        \begin{center}
      \includegraphics[scale=0.39]{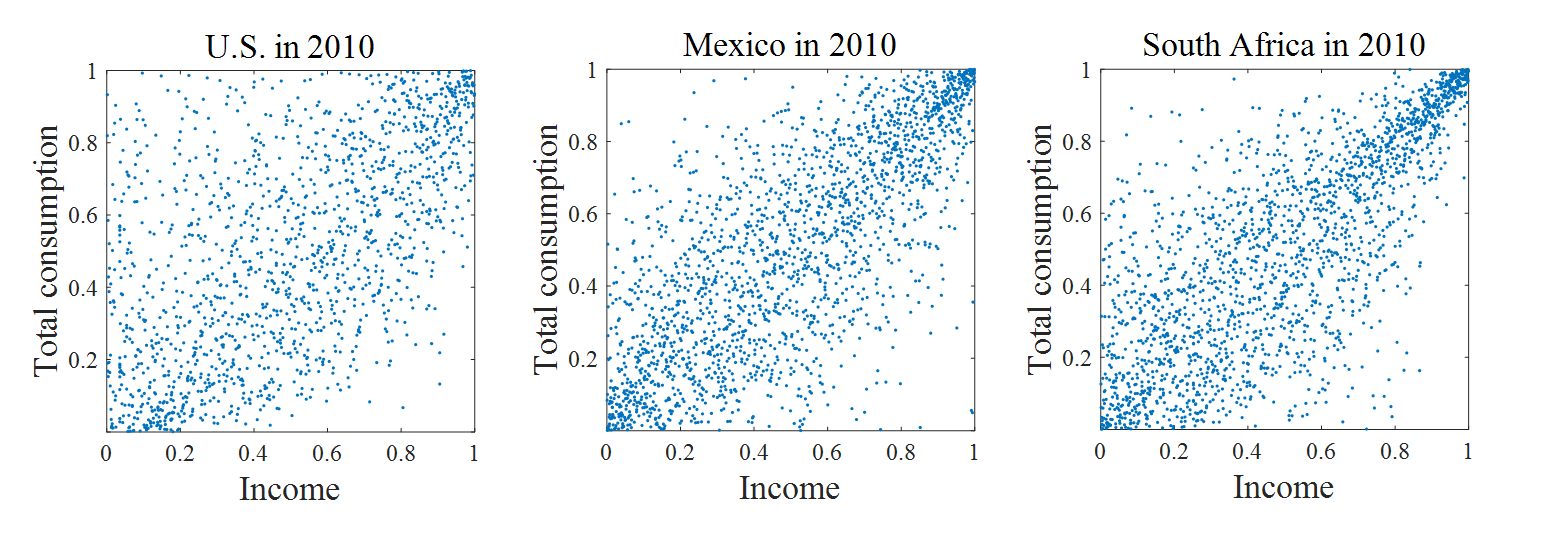}
        \end{center}
      \caption{
The figure displays the scatterplots of household income and consumption in 2010, in the U.S., Mexico, and South Africa after applying the probability integral transforms.
}
\end{figure}

Here, we apply our test to the income and consumption data of the three countries. By applying our test procedure, we may detect difference in the degree of dependence between different group as well as a discrepancy in the structure of the dependence. In the comparison of the U.S., and Mexico, the test statistics are computed as $(T_{n,m}^{(2)}, T_{n,m}^{(\infty )})=(1.49, 3.32)$ and the p-values for both statistics turn out to be zero, suggesting that the dependence structures of income and consumption in the two countries are significantly different. In the comparison of Mexico and South Africa, the test statistics are computed as $(T_{n,m}^{(2)}, T_{n,m}^{(\infty )})=(0.43, 1.30)$. Surprisingly, our randomization test leads to zero p-values for both $T_{n,m}^{(2)}$ and $T_{n,m}^{(\infty )}$ despite that the correlation coefficients and Kendall's $\tau$ are very similar in South Africa and Mexico. This indicates that a strong dissimilarity detected by our test procedure may not be detected by correlation coefficients or Kendall's $\tau$.

\begin{table}[t!]
\begin{center}
\begin{tabular}{c|cc|cc|cc}
\hline\hline
 \multirow{2}{*}{$c$}    & \multicolumn{2}{c}{U.S.}  & \multicolumn{2}{c}{Mexico} & \multicolumn{2}{c}{South Africa}  \\
     \cline{2-7}
  {} & lower $\tau$ & upper $\tau$ & lower $\tau$ & upper $\tau$ & lower $\tau$ & upper $\tau$\\
\hline
0.2&	0.033&	0.192&	0.200&	0.378&	0.128&	0.431\\
0.3&	0.116&	0.205&	0.247&	0.407&	0.178&	0.489\\
0.4&	0.163&	0.240&	0.292&	0.435&	0.213&	0.527\\
0.5&	0.198&	0.264&	0.337&	0.454&	0.248&	0.546\\
0.6&	0.235&	0.282&	0.374&	0.473&	0.291&	0.557\\
0.7&	0.284&	0.316&	0.421&	0.494&	0.344&	0.557\\
0.8&	0.330&	0.359&	0.466&	0.521&	0.409&	0.558\\
\hline\hline
\end{tabular}%
\end{center}
\caption{The table reports the upper and lower conditional Kendall's $\tau$ of income and consumption in the U.S., Mexico and South Africa at exceedance levels $c\in \{0.2,0.3,0.4,0.5,0.6,0.7,0.8\}$. 
}
\end{table}

Table 5.1 provides more detailed information on the structure of income and
consumption dependence in the U.S., Mexico and South Africa in 2010. In the table, we report the
conditional Kendall's $\tau$ of income and consumption at several different exceedance levels.
Recall that for a pair of random variables $(X,Y)$, Kendall's $\tau$ is defined by $\tau (X,Y) =P((X_{i}-X_{j})(Y_{i}-Y_{j})>0)-P((X_{i}-X_{j})(Y_{i}-Y_{j})<0)$, where $(X_{i},Y_{i})$ and $(X_{j},Y_{j})$ are two random draws of $(X,Y)$. Then, the conditional Kendall's $\tau$ at the exceedance level $c\in (0,1)$ is provided by\footnote{The exceedance Kendall's $\tau$ is an analogue to the exceedance correlation in Longin and Solnik (2001), Ang and Chen (2002), and Hong and Zhou (2007). While the exceedance correlation only captures linear conditional dependence, the exceedance Kendall's $\tau$ can also capture nonlinear features of conditional dependence. See also, Manner (2010).}
\begin{eqnarray*}
\tau ^{-}(c) &=&\tau (F_{X}(X),F_{Y}(Y)|F_{X}(X)<c,F_{Y}(Y)<c) \text{\ \ and}\\
\tau ^{+}(c) &=&\tau (F_{X}(X),F_{Y}(Y)|F_{X}(X)>1-c,F_{Y}(Y)>1-c).
\end{eqnarray*}
We observe that in all three countries, the upper conditional dependence is stronger than the lower conditional dependence at any fixed exceedance level. It implies that within a country, the dependence patterns are different depending on relative income and consumption levels. In particular, we found that when both income and consumption levels are relatively low, a greater portion of consumption is on necessities, and the consumption of necessities does not increase much by an increase in the income. On the other hand, when both income and consumption levels are relatively high, consumers spend a greater portion of their income on luxury goods. Hence, the dependence between income and consumption tends to be stronger in the upper tail than in the lower tail.\footnote{Our analysis in Table 5.1 is based on the comparisons between the dependence in the left lower quadrant and right upper quadrant, and the result does not imply that consumption change is more sensitive to income change at higher income level. In fact, the estimated marginal propensity to consumption (MPC) for our data tends to increase when income is low, but it starts to decrease at around the 60th percentile of income in each country. As income increases, the estimated MPC of the U.S. increases
up to 0.32 and decreases to 0.17, while that of Mexico increases 
up to 0.63, and decrease to 0.40. In South Africa, the estimated MPC increases from 0.52 to 0.71, but then decrease again to 0.50.}

In addition, Table 5.1 shows that both the upper conditional dependence and the lower conditional dependence are weaker in the U.S. than in Mexico and South Africa at all exceedance levels. This is consistent with our finding that the U.S. has the lowest correlation coefficient and the lowest Kendall's $\tau$ among the three countries. On the other hand, compared to Mexico, South Africa has smaller lower conditional dependence but larger upper conditional dependence at any fixed exceedance level. As the surplus in upside dependence is offset by the deficit in downside dependence, the value of the correlation coefficients or Kendall's $\tau$ may be similar to that of Mexico. Nevertheless, the patterns of asymmetric dependence in the two countries are very different, and our test procedure can distinguish the dissimilarity of such nonlinear patterns.

\begin{figure}[t!]
  \hspace{-3cm}
        \begin{center}
      \includegraphics[scale=0.39]{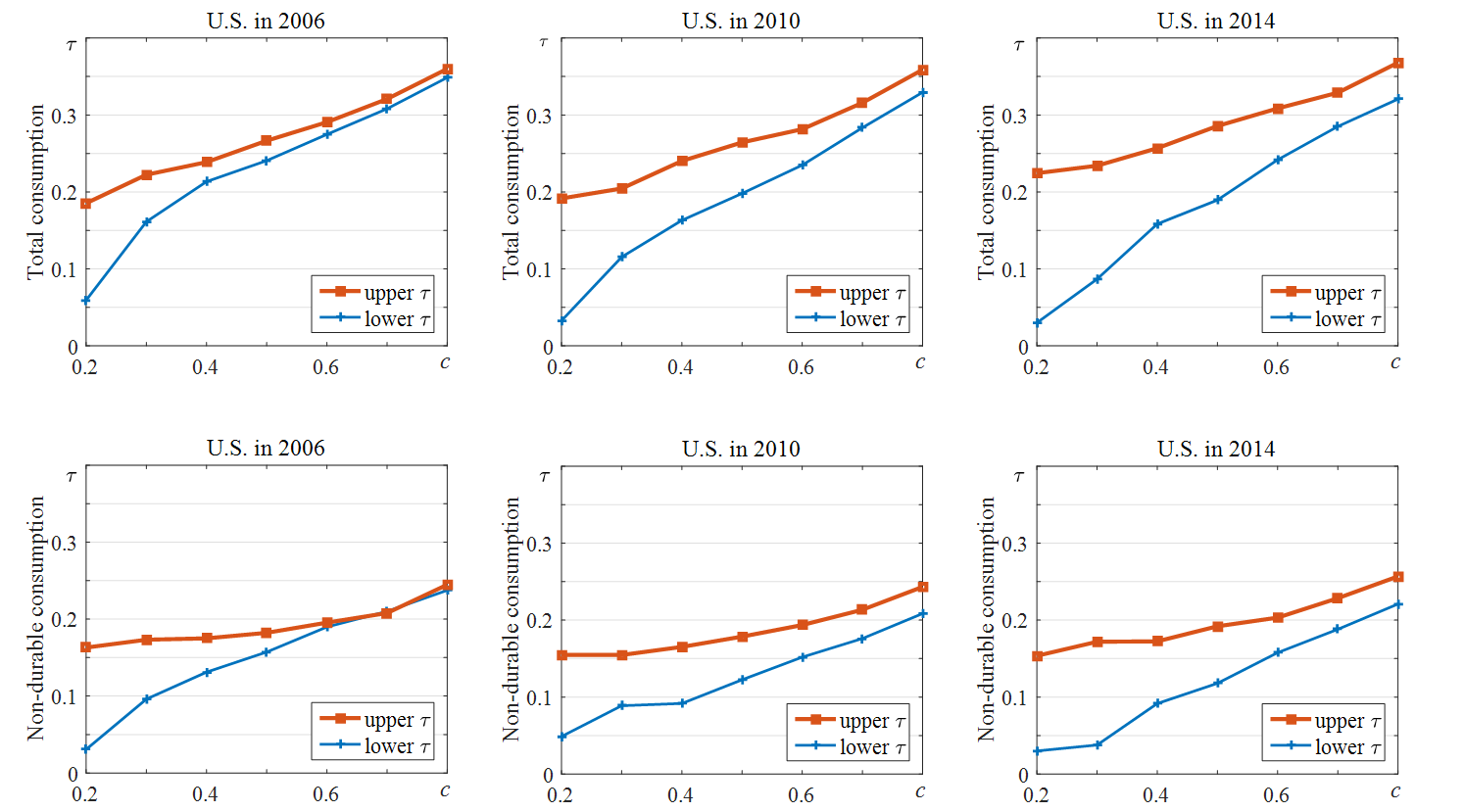}
        \end{center}
      \caption{
We display the graphs of the upper and lower conditional Kendall's $\tau$ of income and consumption in the U.S. in
2006, 2010 and 2014. While the upper conditional dependence does not vary much over the three years, the lower conditional dependence tends to decrease over time.
}
\end{figure}

Since the U.S. Consumer Expenditure Survey provides richer information than the Mexican and South African surveys, we may explore in more depth the recent changes in the income and consumption dependence in the U.S. Hence, we provide additional analysis on the dependence of income and total consumption in the U.S. in 2006, 2010 and 2014, as well as that of income and non-durable consumption in 2006, 2010 and 2014. Following the definition in Attanasio et al.\ (2012), we define the consumption of non-durable goods as the household expenditure on food, clothing, footwear and non-durable entertainment. Based on this definition, we find that the correlation coefficients of income and consumption are very similar in the three years; the correlation coefficients of income and total consumption are 0.5, 0.53 and 0.5 in 2006, 2010 and 2014 respectively, while those of income and non-durable consumption are 0.44, 0.41 and 0.37.

Figure 5.2 provides the graphs of the conditional Kendall's $\tau$ of income and consumption in the U.S. in 2006, 2010 and 2014. For both total consumption and non-durable consumption, the upper conditional Kendall's $\tau$ remains almost the same in the three years, while the lower conditional Kendall's $\tau$ tends to decrease slightly over time. The decrease in the lower conditional dependence is more noticeable between 2006 and 2010 than between 2010 and 2014. Our inference results reflect this finding. In testing for equal dependence of total consumption and income between 2006 and 2010, our p-values for $T_{n,m}^{(2)}$ and $T_{n,m}^{(\infty )}$ are computed as 0.04 and 0.08, while they are 0.30 and 0.99 between 2010 and 2014. Using non-durable consumption, we obtain slightly smaller p-values. The p-values for $T_{n,m}^{(2)}$ and $T_{n,m}^{(\infty )}$ are 0.03 and 0.01 between 2006 and 2010, while they are 0.22 and 0.94 between 2010 and 2014. The results suggest that for both specifications of consumption, the dependence structure of income and consumption is significantly different between 2006 and 2010 at the $5\%$ and $10\%$ significance levels. However, the difference is less significant between 2010 and 2014.

\subsection{Brexit effect on financial integration in Europe}
Since established, the European Union (EU) has served as a coalition
which fostered the integration of economic development among European
countries. The economic integration has forced the economic decisions of the
countries in the EU to be intimately dependent on each other. The countries
have benefited from the upside gains but also suffered from the downside
losses. Among others, the U.K. has played a central role in
the decision making process of EU, providing a large portion of its budget over the past years. However, in pursuit of financial and political
independence, the U.K. eventually announced `Brexit' (Britain's exit from the
EU) in June of 2016. The term `Brexit', which had been previously used for the
potential exit of the U.K. from the EU, does not refer to a hypothetical
situation anymore.

The announcement of Brexit created a huge external shock to
the international economy. Following the announcement, there have been
massive debates on what would happen to the U.K. and EU in the future.
One of the major concerns is the impact of the U.K.'s decision on the economic integration in Europe.
In this section, we aim to provide statistical evidence of the `Brexit effect' on European financial market integration by investigating stock returns in the U.K., France, Germany and Switzerland.
Although the process of Brexit is still ongoing and no definitive answer may
be given to the question of the long run effect of Brexit, our findings in
this section will provide evidence of the Brexit effect in its early
and transitional stage.

\begin{figure}[t!]
  \hspace{-3cm}
          \begin{center}
      \includegraphics[scale=0.35]{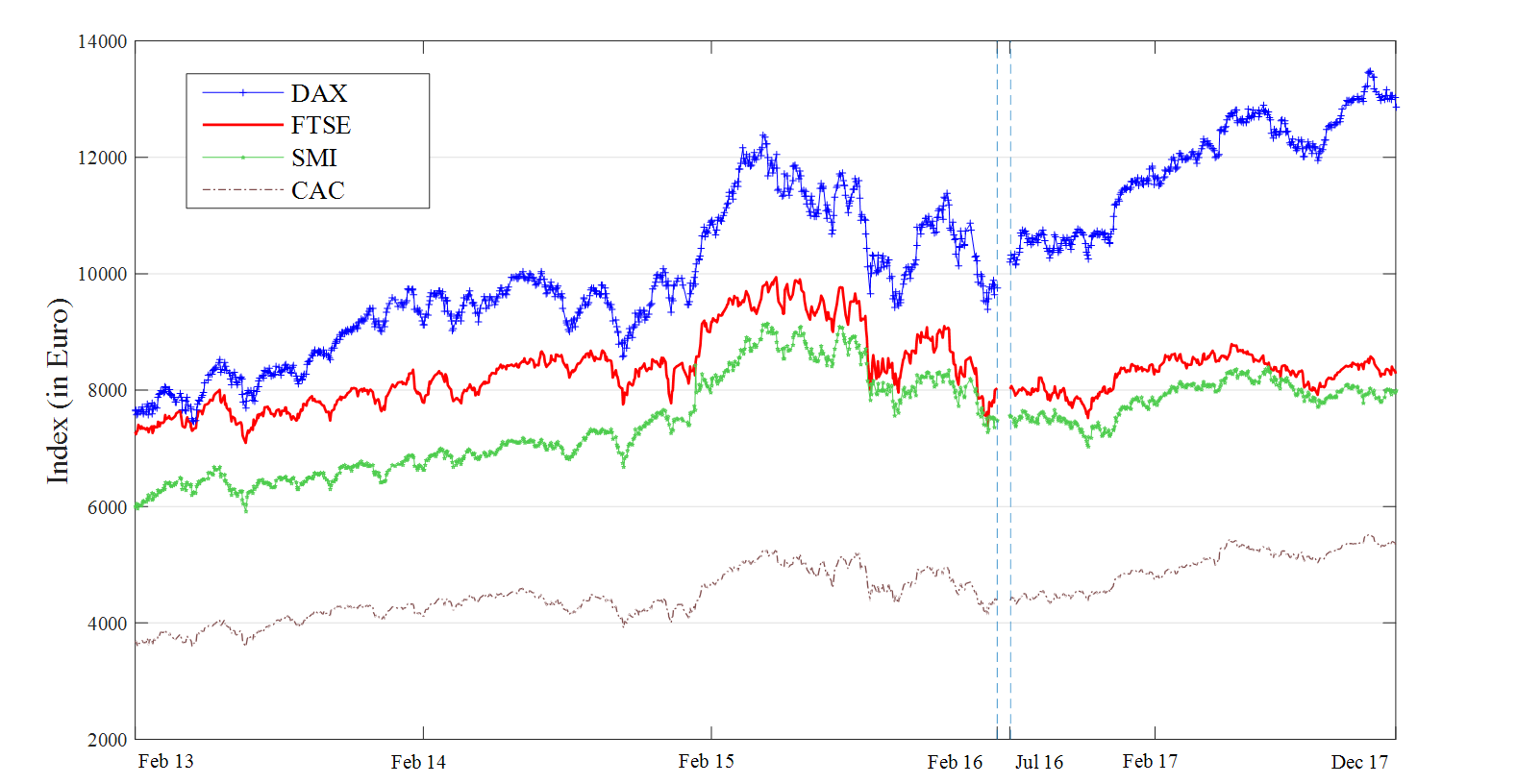}
        \end{center}
      \caption{
We display time plots of FTSE, CAC, DAX and SMI indices in the pre-Brexit and Brexit periods. The stock indices have a strong tendency to move together in both pre-Brexit and Brexit periods.
      }
\end{figure}

For our analysis, we collect the four stock indices of FTSE (U.K.), CAC
(France), DAX (Germany) and SMI (Switzerland) from February 2013 to July 2017,
at a daily frequency. Using the announcement of the Brexit as a
cutoff point, we divide our observations into two groups, the data of
`before' the Brexit announcement and `after' the Brexit announcement. To be
more precise, we define the `pre-Brexit' period as February
2013 through January 2016, and `Brexit' period as July 2016 through December 2017. Although the referendum for Brexit was held in June 2016,
the financial market structure in Europe has been potentially influenced by
the announcement in February 2016 of making such a poll. Thus, we leave out the
observations from February 2016 to June 2016 in defining the `pre-Brexit' period.%
\footnote{%
We have also examined the Brexit effect on the European financial market
based on different definitions of pre-Brexit period. Neither of using one
year nor two years of observations before February 2016 as pre-Brexit data changes
the overall inference results presented in this section.} After excluding weekends and public holidays, the sample sizes are 734 for the pre-Brexit period and 338 for the Brexit period.
These indices, as displayed in Figure 5.3, present a high degree of comovement in both pre-Brexit and Brexit periods.
To remove temporal dependence, we apply the AR(1)-GJR-GARCH $(1,1)$ filter to the stock returns obtained by taking the log-difference of the indices.\footnote{%
We examined several different GARCH specifications with student $t$,
skewed $t$ and Gaussian innovations, and found that the AR(1)-GJR-GARCH(1,1) model with student $t$ innovations provides the best fit.
After applying a proper GARCH filter, the filtered stock returns are conventionally regarded as i.i.d. in the empirical finance literature.
See Abhyankar et al. (1997), Manner(2001), Hu (2006), Roch and Alegre (2006), Cotter (2007), Aas et al. (2009), Giacomini et al. (2009), Hu and Kercheval (2010), Aloui et al. (2011), Nikoloulopoulos et al. (2012), St\o{}ve et al. (2014) and many others. Specifically, Chan et al. (2009) provided a theoretical justification for the GARCH residuals based estimation of copulas in the semi-parametric setting.}

\begin{figure}[t!]
  \hspace{-3cm}
        \begin{center}
      \includegraphics[scale=0.39]{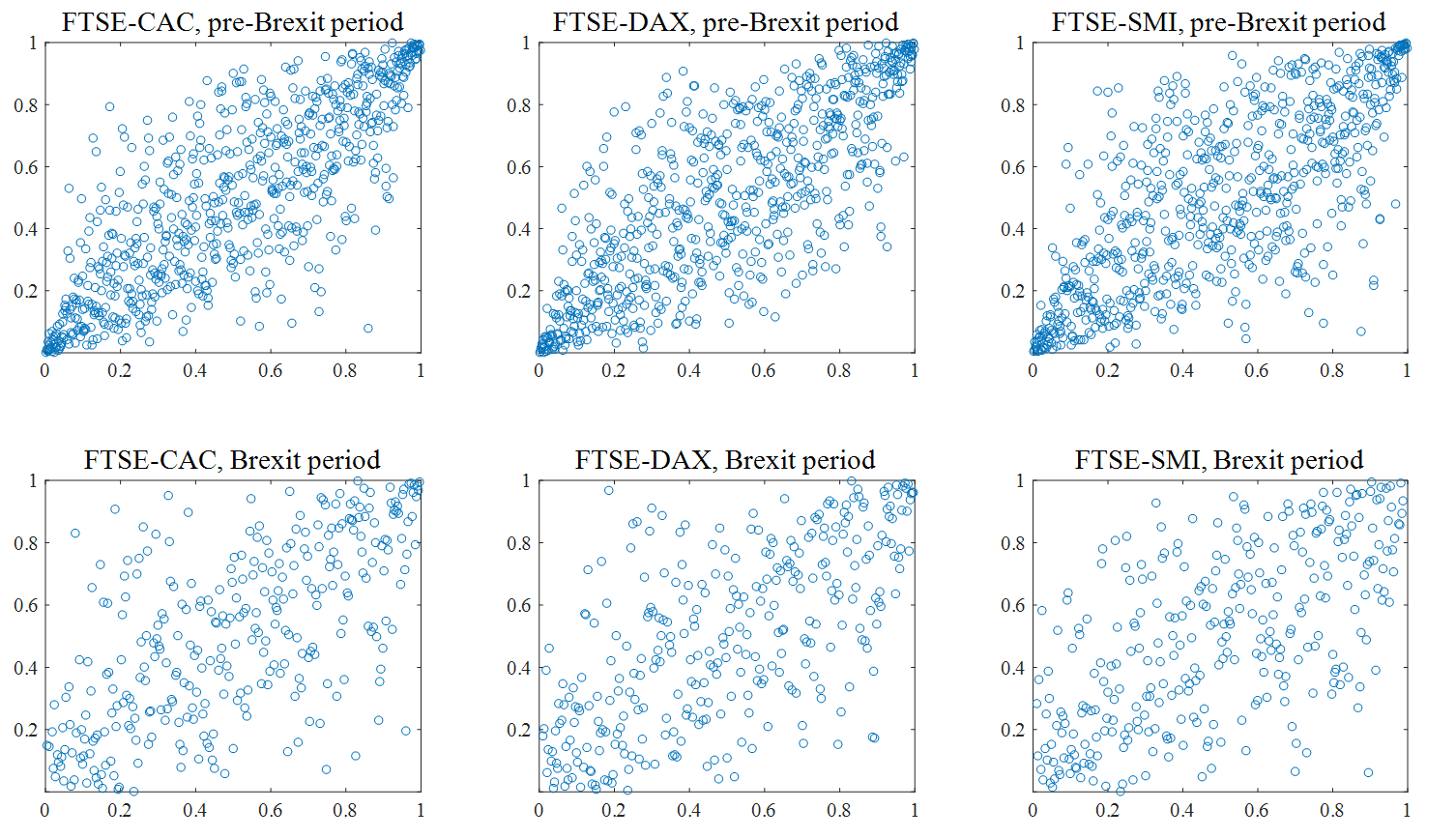}
        \end{center}
      \caption{
We display the scatterplots of the stock returns of FTSE against CAC, DAX, and SMI after applying the probability integral transformation. The three plots on the top illustrate the bivariate dependencies of FTSE and other stock returns in the pre-Brexit period and the three plots at the bottom illustrate those in the Brexit period.
}
\end{figure}

Figure 5.4 displays scatterplots of the FTSE against CAC, DAX, and
SMI in the pre-Brexit period and the Brexit period. The stock returns have been normalized by applying the probability integral transforms. Casual inspection reveals
that the observations are less concentrated on the diagonal line in the Brexit period, suggesting that the cross market comovement of stock returns
has become weaker after the U.K. decided to leave the EU. This can also be
observed from the changes in the correlation coefficient. The correlation
coefficient of FTSE and CAC is 0.86 in the pre-Brexit period and 0.71 in
the Brexit period. Similarly, the correlation coefficients of FTSE and DAX, and of FTSE and SMI also have decreased from 0.82 to 0.65 and from 0.71 to 0.66,
respectively.

\begin{table}[t!]
\begin{center}
\begin{tabular}{l||lc|lc}
\hline\hline
stock returns & $T_{n,m}^{(2)}$ & p-value & $T_{n,m}^{(\infty)}$ & p-value \\ \hline\hline
(FTSE, CAC) &0.253**	&0.000	&0.862**	&0.000 \\
(FTSE, DAX) &0.239**	&0.000	&0.731**	&0.024 \\
(FTSE, SMI) &0.143	&0.154	&0.599	&0.234 \\ \hline
(CAC, DAX)  &0.124**	&0.005	&0.615**	&0.048 \\
(CAC, SMI)  &0.144*	&0.092	&0.584	&0.262 \\
(DAX, SMI)  &0.134	&0.190	&0.591	&0.252 \\ \hline\hline
(CAC, DAX, SMI) &0.168**	&0.048	&0.721	&0.237 \\
(FTSE, CAC, DAX, SMI) &0.299**	&0.002	&1.036**	&0.021 \\ \hline\hline
\end{tabular}%
\end{center}
\caption{The table shows the test statistics and p-values of our randomization test for equality in dependence structure. The null hypotheses concern the six pairwise dependence of (FTSE, CAC), (FTSE, DAX), (FTSE, SMI), (CAC, DAX), (CAC, SMI), (DAX, SMI), and the mutual dependence of (CAC, DAX, SMI), (FTSE, CAC, DAX, SMI). The critical values are obtained from 1000 random permutations to test each hypothesis.
}
\end{table}

In Table 5.2, we report our randomization test results concerning the Brexit effect. Firstly, our results show that the U.K.'s financial market dependence on the French and German financial markets has completely changed after
the Brexit announcement. Applying our randomization test to the pairs of
stock returns (FTSE, CAC) and (FTSE, DAX) yields p-values close to
zero, which indicates overwhelming rejection of equality in the pairwise
dependence. This change is due to the overall decrease in dependency (a change in the degree of dependence), and unequal decrease in dependency--more decrease in the downside than in the upside-- that leads to asymmetric dependence (a change in the structure of dependence).
For a fixed level $c=0.5$ for instance, the upper conditional Kendall's $\tau$ of FTSE and CAC slightly decreased from $0.76$ to $0.65$ while the lower conditional Kendall's $\tau$ substantially decreased from $0.77$ to $0.26$. In the (FTSE, DAX) pair, the upper conditional Kendall's $\tau$ at level $c=0.5$ decreased from $0.44$ to $0.35$, while the lower conditional Kendall's $\tau$ decreased from $0.43$ to $0.14$, and again, the decrease is more prominent in the downside than in the upside.
On the other hand, the change of the dependence structure in the pair of (FTSE, SMI) turns out
to be less significant: we fail to reject the null hypothesis of copula equality at both $5\%$ and $10\%$
significance levels. We also found that in terms of distributional symmetry, no obvious change in the pair (FTSE, SMI) is observed after the Brexit announcement.

While there is no universal consensus on the extension of the correlation coefficient or Kendall's $\tau$ to more than two variables, our statistical procedure can be
naturally applied to study higher dimensional dependence structures. For
instance, by comparing the four dimensional copulas of stock returns to the
FTSE, CAC, DAX, and SMI in the pre-Brexit and Brexit period, we can perform a statistical test to examine the Brexit effect on the dependence structure of the four returns.
It turns out that our randomization tests with the statistics $T_{n,m}^{(2)}$ and $T_{n,m}^{(\infty )}$ yield p-values of $0.01$ and
$0.02$ respectively, indicating that there has been a significance change in mutual dependence of the stock returns of FTSE, CAC, DAX, and SMI after the announcement of Brexit.
Similarly, we can also examine the Brexit effect on the dependence of CAC, DAX and SMI, excluding FTSE. The question arises when our concern is to test for the Brexit effect on the financial market dependence among remaining EU countries. Here, our inference provides somewhat mixed results: we detect a significant change in the
dependence structure of CAC, DAX and SMI with the test statistic $T_{n,m}^{(2)}$ at both $5\%$ and $10\%$ levels, but the
change is less significant with $T_{n,m}^{(\infty )}$.

To summarize, our empirical findings reveal that the European financial market has experienced a substantial change after the Brexit announcement. Firstly, the dependence of financial market returns between the U.K. and other European countries has decreased after the Brexit announcement. In particular, we observe a greater decrease in dependence during market downturns than market upturns, which implies that the U.K.'s financial market is now less likely to crash together with other financial markets. Secondly, the financial integration among the U.K., France, Germany and Switzerland has decreased in reaction to the announcement of Brexit. Our test results indicate that there has been a significant change in the dependence structure of the stock markets in the four countries. We also found some evidence that the announcement of Brexit not only changed the financial market dependence of the U.K., but also that of the remaining EU countries. Lastly, the change in the dependence structure seems to be more recognizable among the larger economies. In our analysis, Brexit has caused more significant influence on financial market dependence among the U.K., France and Germany, while Switzerland has been less affected by the Brexit announcement.


\section{Final remarks}
Randomization tests provide useful tools for inference in situations where the null hypothesis implies that the distribution of the data is invariant to a group of transformations.
In particular, permutation tests have been widely applied for testing the equality of distribution functions.
Since copulas constitute a class of distribution functions, they may be naturally employed for testing the equality of copulas.
However, the problem has not been examined
to date and this is an important missing point in the literature.

This paper examines the use of permutation methods for testing copula equality. Unfortunately, the classical permutation
method is not applicable because we do not observe samples directly from the
copulas. Although we may instead explore the pseudo samples which consist of normalized ranks, the
application of the permutation method still requires caution due to the
distortion that permutation causes to the univariate margins of the pseudo
samples. Asymptotically valid inference can be achieved by either modifying the permutation method considering the additional terms of the limit distribution induced by the uncertainty in margins (Theorem 3.1), or by correcting the distortion through taking one more step of normalizing the margins after each permutation (Theorem 3.2).

\newpage
\appendix
\section{Proofs and some numerical results}

\textit{\textbf{Lemma A.1. }}When $C_{1}=C_{2}$, the law of $\mathbb{\tilde{T%
}}\equiv \sqrt{1-\lambda }\mathbb{B}_{\bar{C}}-\sqrt{\lambda }\mathbb{B}_{%
\bar{C}}^{\prime }$ is equivalent to the law of $\mathbb{\tilde{T}}%
_{0}\equiv \sqrt{1-\lambda }\mathbb{B}_{C_{1}}-\sqrt{\lambda }\mathbb{B}%
_{C_{2}}$.

\bigskip
\textbf{Proof of Lemma A.1. } Since $\mathbb{\tilde{T}}$ and $%
\mathbb{\tilde{T}}_{0}$ are centered Gaussian processes, it suffices to show
that their covariance structures are equivalent. We will prove it by showing that under the condition of $C_{1}=C_{2}$, each covariance kernel of $\mathbb{\tilde{T}}$ and $\mathbb{\tilde{T}}_{0}$ reduces to that of $\mathbb{B}_{C_{1}}$. Recall that $\mathbb{B}_{C_{1}}$ is a
Brownian bridge on $[0,1]^{d}$ with covariance kernel
\begin{equation*}
\text{Cov}\left( \mathbb{B}_{C_{1}}(\mathbf{u}),\mathbb{B}_{C_{1}}(\mathbf{u}%
^{\prime })\right) =C_{1}(\mathbf{u}\wedge \mathbf{u}^{\prime })-C_{1}(%
\mathbf{u})C_{1}(\mathbf{u}^{\prime })
\end{equation*}%
for $\mathbf{u}$ and $\mathbf{u}^{\prime }$ in $[0,1]^{d}$.

Firstly, since $\mathbb{B}_{C_{1}}$ and $\mathbb{B}_{C_{2}}$ are independent, the covariance kernel of $\mathbb{\tilde{T}}_{0}$ can be written as,
\begin{eqnarray*}
\text{Cov}\left( \mathbb{\tilde{T}}_{0}(\mathbf{u}),\mathbb{\tilde{T}}_{0}(%
\mathbf{u}^{\prime })\right)  &=&E\left\{ \left( \sqrt{1-\lambda }\mathbb{B}%
_{C_{1}}(\mathbf{u})-\sqrt{\lambda }\mathbb{B}_{C_{2}}(\mathbf{u})\right)
\left( \sqrt{1-\lambda }\mathbb{B}_{C_{1}}(\mathbf{u}^{\prime })-\sqrt{%
\lambda }\mathbb{B}_{C_{2}}(\mathbf{u}^{\prime })\right) \right\}  \\
&=&(1-\lambda )\left( C_{1}(\mathbf{u}\wedge \mathbf{u}^{\prime })-C_{1}(%
\mathbf{u})C_{1}(\mathbf{u}^{\prime })\right) -\lambda \left( C_{2}(\mathbf{u%
}\wedge \mathbf{u}^{\prime })-C_{2}(\mathbf{u})C_{2}(\mathbf{u}^{\prime
})\right)  \\
&=&C_{1}(\mathbf{u}\wedge \mathbf{u}^{\prime })-C_{1}(\mathbf{u})C_{1}(%
\mathbf{u}^{\prime })
\end{eqnarray*}%
where the last equality follows from the condition $C_{1}=C_{2}$.

Second, using the independence of $\mathbb{B}_{\bar{C}}$ and $\mathbb{B}_{\bar{C}}^{\prime }$, the covariance kernel of $\mathbb{\tilde{T}}$ can be also written as,
\begin{eqnarray*}
\text{Cov}\left( \mathbb{\tilde{T}}(\mathbf{u}),\mathbb{\tilde{T}}(\mathbf{u}%
^{\prime })\right)  &=&E\left\{ \left( \sqrt{1-\lambda }\mathbb{B}_{\bar{C}}(%
\mathbf{u})-\sqrt{\lambda }\mathbb{B}_{\bar{C}^{\prime }}(\mathbf{u})\right)
\left( \sqrt{1-\lambda }\mathbb{B}_{\bar{C}}(\mathbf{u}^{\prime })-\sqrt{%
\lambda }\mathbb{B}_{\bar{C}^{\prime }}(\mathbf{u}^{\prime })\right)
\right\}  \\
&=&(1-\lambda )(\bar{C}(\mathbf{u}\wedge \mathbf{u}^{\prime })-\bar{C}(%
\mathbf{u})\bar{C}(\mathbf{u}^{\prime }))-\lambda (\bar{C}(\mathbf{u}\wedge
\mathbf{u}^{\prime })-\bar{C}(\mathbf{u})\bar{C}(\mathbf{u}^{\prime })) \\
&=&\bar{C}(\mathbf{u}\wedge \mathbf{u}^{\prime })-\bar{C}(\mathbf{u})\bar{C}(%
\mathbf{u}^{\prime })\text{\ \ \ \ \ \ } \\
&=&C_{1}(\mathbf{u}\wedge \mathbf{u}^{\prime })-C_{1}(\mathbf{u})C_{1}(%
\mathbf{u}^{\prime })
\end{eqnarray*}%
The last equality is from the condition that $C_{1}=C_{2}=\bar{C}$ (see Remark 3.1). This completes our proof.

\bigskip \textbf{Proof of Lemma 3.1. }The proof is straightforward from an application of
the continuous mapping theorem to the weak convergence of the empirical copula process.

\bigskip \textbf{Proof of Lemma 3.2. } For any $\mathbf{u}%
=(u_{1},...,u_{d})\in \lbrack 0,1]^{d}$, $\tilde{C}_{1,n}$ is linear in the
sense that we may write%
\begin{equation*}
\sqrt{n}(\tilde{C}_{1,n}(Z)(\mathbf{u})-C_{1}(\mathbf{u}))=\frac{1}{\sqrt{n}}%
\sum_{i=1}^{n}\left\{ 1(U^{i}\leq \mathbf{u})-E(1(U^{i}\leq \mathbf{u}%
))\right\} \text{.\ }
\end{equation*}%
Note that $\sqrt{n}(\tilde{C}_{1,n}(Z)-C_{1})$ is simply the standard empirical
process, and the limit of $\sqrt{n}(\tilde{C}_{1,n}(Z)-C_{1})$ is the $C_{1}$%
-Brownian bridge, $\mathbb{B}_{C_{1}}$. The same point can be made with
regard to any properly defined copula. Therefore, the empirical processes
associated with $C_{2}$ and $\bar{C}$ are also linear and their limits are\ $%
\mathbb{B}_{C_{2}}$ and $\mathbb{B}_{\bar{C}}$, respectively. Then the
remaining proof of Lemma 3.2 can be done by verifying the key conditions in
Chung and Romano (2013).

Let $Q_{1}^{i}(Z)(\mathbf{u})=1(U^{i}\leq \mathbf{u})$ and $Q_{2}^{j}(Z)(\mathbf{u})=1(V^{j}\leq \mathbf{u}%
)$ for $i=1,...,n$ and $j=1,...,m$, and let $\bar{Q}_{1,n}$ and $\bar{Q}_{2,m}$ be the sample means
calculated from $\{Q_{1}^{i}\}_{i=1}^{n}$ and $\{Q_{2}^{j}\}_{j=1}^{m}$
respectively. We observe that $\tilde{T}_{n,m}(Z)$ is simply a scaled difference
between the two sample means,
\begin{equation*}
\tilde{T}_{n,m}(Z)=\sqrt{\frac{nm}{n+m}}\left( \bar{Q}_{1,n}(Z)-\bar{Q}%
_{2,m}(Z)\right) \text{,}
\end{equation*}
and the limit of its sampling distribution is $\sqrt{1-\lambda }%
\mathbb{B}_{C_{1}}-\sqrt{\lambda }\mathbb{B}_{C_{2}}$ under $C_{1}=C_{2}$.
On the other hand, the limit of its permutation distribution can be inferred from its asymptotic unconditional distribution when
the samples are drawn from that mixture distribution. Since the mixture distribution of $U=(U_{1},...,U_{d})$ and $V=(V_{1},...,V_{d})$ is the mixture of the copula of $U$ and $V$, i.e., $\bar{C}=\lambda C_{1}+(1-\lambda )C_{2}$,  by Chung and Romano (2013) we have
\begin{equation*}
(\tilde{T}_{n,m}(Z^{\pi }),\tilde{T}_{n,m}(Z^{\pi ^{\prime
}}))\rightsquigarrow (\mathbb{\tilde{T}},\mathbb{\tilde{T}}^{\prime })
\end{equation*}
where $\mathbb{\tilde{T}}$ is $\sqrt{1-\lambda }\mathbb{B}_{\bar{C}}-\sqrt{\lambda
}\mathbb{B}_{\bar{C}}^{\prime }$ and $\mathbb{\tilde{T}}^{\prime}$ is an independent copy of $\mathbb{\tilde{T}}$. Note that the conditions (5.9) and (5.10) in Chung and Romano (2013) are trivially satisfied because this is simply the problem of the difference in means. The result is also consistent with Lemma 4.3 in Romano (1990) and
Theorem 1 in Wu (1990).

\bigskip \textbf{Proof of Lemma 3.3. }We start by noting that the specific
form of $1(\hat{Z}^{i}\leq \mathbf{u})$ is determined by the range of $i$; $%
1(\hat{Z}^{i}\leq \mathbf{u})$ is $1(\hat{U}_{n}^{i}\leq \mathbf{u}) $ for $%
1\leq i\leq n$, and $1(\hat{V}_{m}^{i-n}\leq \mathbf{u})$ for $n<i \leq N$. Motivated
from this observation, we consider the following decomposition of $\tilde{T}%
_{n,m}(\hat{Z}^{\pi })$:
\begin{eqnarray*}
\tilde{T}_{n,m}(\hat{Z}^{\pi })(\mathbf{u}) &=&\sqrt{\frac{nm}{n+m}}\left\{
\left( \frac{1}{n}\sum_{i=1}^{n}1(\hat{Z}^{\pi (i)}\leq \mathbf{u},\pi
(i)\leq n)+\frac{1}{n}\sum_{i=1}^{n}1(\hat{Z}^{\pi (i)}\leq \mathbf{u},\pi
(i)>n)\right) \right. \\
&&\left. -\left( \frac{1}{m}\sum_{i=n+1}^{N}1(\hat{Z}^{\pi (i)}\leq \mathbf{u%
},\pi (i)\leq n)+\frac{1}{m}\sum_{i=n+1}^{N}1(\hat{Z}^{\pi (i)}\leq \mathbf{u%
},\pi (i)>n)\right) \right\} \\
&=&\sqrt{\frac{nm}{n+m}}\left\{ \left( \frac{1}{n}\sum_{i=1}^{n}1(\hat{Z}%
^{\pi (i)}\leq \mathbf{u},\pi (i)\leq n)-\frac{1}{m}\sum_{i=n+1}^{N}1(\hat{Z}%
^{\pi (i)}\leq \mathbf{u},\pi (i)\leq n)\right) \right. \\
&&\left. +\left( \frac{1}{n}\sum_{i=1}^{n}1(\hat{Z}^{\pi (i)}\leq \mathbf{u}%
,\pi (i)>n)-\frac{1}{m}\sum_{i=n+1}^{N}1(\hat{Z}^{\pi (i)}\leq \mathbf{u}%
,\pi (i)>n)\right) \right\} \text{. }
\end{eqnarray*}
The first two terms in the last equation collect the terms of the form $1(%
\hat{U}_{n}^{i}\leq u)$ for $i=1,...,n$, and the last two terms in the last equation collect the terms
of the form $1(\hat{V}_{m}^{j}\leq u)$ for $j=1,...,m$.

Now for $\pi \in \mathbf{G}_{N}$, define $\alpha _{n,m}^{\pi }(\mathbf{u})$ and $\beta _{n,m}^{\pi }(\mathbf{u})$ as
\begin{equation*}
\alpha _{n,m}^{\pi }(\mathbf{u})=\sqrt{\frac{nm}{n+m}}\left( \frac{1}{n}%
\sum_{i=1}^{n}1(Z^{\pi (i)}\leq \mathbf{u},\pi (i)\leq n)-\frac{1}{m}%
\sum_{i=n+1}^{N}1(Z^{\pi (i)}\leq \mathbf{u},\pi (i)\leq n)\right)
\end{equation*}%
and%
\begin{equation*}
\beta _{n,m}^{\pi }(\mathbf{u})=\sqrt{\frac{nm}{n+m}}\left( \frac{1}{n}%
\sum_{i=1}^{n}1(Z^{\pi (i)}\leq \mathbf{u},\pi (i)>n)-\frac{1}{m}%
\sum_{i=n+1}^{N}1(Z^{\pi (i)}\leq \mathbf{u},\pi (i)>n)\right) \text{. }
\end{equation*}%
Using the notations $\alpha _{n,m}^{\pi }$ and $\beta _{n,m}^{\pi }$, we can rewrite $%
\tilde{T}_{n,m}(Z^{\pi })$ and $\tilde{T}_{n,m}(\hat{Z}^{\pi })$ in the following way. Let $D_{n}(\mathbf{u}%
)=(D_{1,n}(u_{1}),...,D_{d,n}(u_{d}))$ and $E_{m}(\mathbf{u}%
)=(E_{1,m}(u_{1}),...,E_{d,m}(u_{d}))$, where those components are defined by $D_{q,n}(u_{q})=\frac{1}{n}%
\sum_{i=1}^{n}1(F_{q}(X_{q}^{i})\leq u_{q})$ and $E_{q,m}(u_{q})=\frac{1}{m}%
\sum_{j=1}^{m}1(G_{q}(Y_{q}^{j})\leq u_{q})$.
Let $D_{n}^{-1}$ and $E_{m}^{-1}$ be the componentwise generalized inverse, as in R\'{e}millard and Scaillet (2009; 383p). Then, we have
\begin{eqnarray*}
\tilde{T}_{n,m}(Z^{\pi })(\mathbf{u}) &=&\alpha _{n,m}^{\pi }(\mathbf{u}%
)+\beta _{n,m}^{\pi }(\mathbf{u})\text{ \ and} \\
\tilde{T}_{n,m}(\hat{Z}^{\pi })(\mathbf{u}) &=&\alpha _{n,m}^{\pi
}(D_{n}^{-1}(\mathbf{u}))+\beta _{n,m}^{\pi }(E_{m}^{-1}(\mathbf{u})).
\end{eqnarray*}

The joint convergence of $\alpha _{n,m}^{\pi }$ and $\beta _{n,m}^{\pi }$
can be shown using the same argument in Lehmann and Romano (2005; Example
15.2.6 and Theorem 15.2.5). To understand how, it helps to see how $\alpha _{n,m}^{\pi }$ and $\beta _{n,m}^{\pi }$ can be reformulated in their framework. The terms $\alpha _{n,m}^{\pi }$ and $\beta _{n,m}^{\pi }$ can be written in terms of $Q_{1}^{i}$ and $Q_{2}^{j}$ defined in our proof of Lemma 3.2. Let $%
\Gamma=(\Gamma ^{1},\Gamma ^{2},...,\Gamma
^{N})=(Q_{1}^{1},...,Q_{1}^{n},0,...,0)$ and let $\Upsilon=(\Upsilon
^{1},\Upsilon ^{2},...,\Upsilon ^{N})=(0,...,0,Q_{2}^{1},...,Q_{2}^{m})$.
We have,
\begin{equation*}
(\alpha _{n,m}^{\pi },\beta _{n,m}^{\pi })=\left( \sqrt{\frac{nm}{n+m}}%
\sum_{i=1}^{N}\Gamma ^{i}\kappa ^{\pi (i)},\sqrt{\frac{nm}{n+m}}%
\sum_{i=1}^{N}\Upsilon ^{i}\kappa ^{\pi (i)}\right)
\end{equation*}%
where $\kappa ^{\pi (i)}$ is defined to be $\frac{1}{n}$ when $\pi (i)\leq n$
and $-\frac{1}{m}$ otherwise.

Now, let the individual limits of $\alpha
_{n,m}^{\pi }$ and $\beta _{n,m}^{\pi }$ be $\alpha ^{\ast }$ and $\beta
^{\ast }$ respectively. The triangular inequality implies that
\begin{eqnarray*}
&&\sup_{\mathbf{u}\in \lbrack 0,1]^{d}}\left\vert \alpha _{n,m}^{\pi
}(D_{n}^{-1}(\mathbf{u}))+\beta _{n,m}^{\pi }(E_{m}^{-1}(\mathbf{u}%
))-(\alpha ^{\ast }(\mathbf{u})+\beta ^{\ast }(\mathbf{u}))\right\vert  \\
&\leq &\sup_{\mathbf{u}\in \lbrack 0,1]^{d}}\left\vert \alpha _{n,m}^{\pi
}(D_{n}^{-1}(\mathbf{u}))-\alpha ^{\ast }(\mathbf{u})\right\vert +\sup_{%
\mathbf{u}\in \lbrack 0,1]^{d}}\left\vert \beta _{n,m}^{\pi }(E_{m}^{-1}(%
\mathbf{u}))-\beta ^{\ast }(\mathbf{u})\right\vert
\end{eqnarray*}%
where $\alpha ^{\ast }+\beta ^{\ast }$ is $\mathbb{\tilde{T}}$. Since $\sup_{%
\mathbf{u}\in \lbrack 0,1]^{d}}\left\vert D_{n}^{-1}(\mathbf{u})-\mathbf{u}%
\right\vert \rightarrow 0$ and $\sup_{\mathbf{u}\in \lbrack
0,1]^{d}}\left\vert E_{m}^{-1}(\mathbf{u})-\mathbf{u}\right\vert \rightarrow
0$ as $n$ and $m\rightarrow \infty $ (Shorack and Wellner, 1986; R\'{e}%
millard and Scaillet, 2009), we conclude $\tilde{T}_{n,m}(\hat{Z}^{\pi })\rightsquigarrow \mathbb{\tilde{T}}$\text{.}

Lastly, consider a permutation $\pi ^{\prime }\in \mathbf{G}_{N}$ that is independent of $\pi$.
Let $\hat{Q}%
^{i}=1(\hat{Z}^{i}\leq \mathbf{u})$ for $i=1,...,N$ and $\kappa ^{\pi
^{\prime }(i)}$ be $\frac{1}{n}$ if $\pi ^{\prime }(i)\leq n$ and $-\frac{1}{%
m}$ otherwise. Then, we see that
\begin{equation*}
Cov(\tilde{T}_{n,m}(\hat{Z}^{\pi }),\tilde{T}_{n,m}(\hat{Z}^{\pi ^{\prime
}}))=\sqrt{\frac{nm}{n+m}}Cov\left( \sum_{i=1}^{N}\hat{Q}^{i}\kappa ^{\pi
(i)},\sum_{i=1}^{N}\hat{Q}^{i}\kappa ^{\pi ^{\prime }(i)}\right) =0\text{.}
\end{equation*}%
The independence of $\mathbb{\tilde{T}}$ and $\mathbb{\tilde{T}}^{\prime }$
follows from the independence of $\pi $ and $\pi ^{\prime }$, as in Lehmann
and Romano (2005; 641p-642p).

\bigskip \textbf{Proof of Theorem 3.1.} We invoke the Slutsky's theorem for
randomization distributions, treating copula derivatives as non-random constants for each $\mathbf{u}$.
Let $\mathbb{C}_{0}=\sqrt{%
1-\lambda }\mathbb{C}_{C_{1}}-\sqrt{\lambda }\mathbb{C}_{C_{2}}$ and $%
\mathbb{C}_{0}^{\prime }$ be an independent copy of $\mathbb{C}_{0}$. By
Lemma A.1, Lemma 3.3, Theorem 5.1 and Theorem 5.2 in Chung and Romano
(2013), we have,%
\begin{equation*}
(\mathfrak{T}_{n,m}^{\pi },\mathfrak{T}_{n,m}^{\pi ^{\prime
}})\rightsquigarrow (\mathbb{C}_{0},\mathbb{C}_{0}^{\prime })
\end{equation*}%
whenever $C_{1}=C_{2}$. In terms of conditional convergence, we have $\mathfrak{%
T}_{n,m}^{\pi }\overset{\mathrm{P}}{\underset{\pi }{\rightsquigarrow }}%
\mathbb{C}_{0}$ under the null hypothesis, and our proof is done by applying the continuous mapping
theorem for the conditional convergence (Lemma A.2 below),
\begin{equation*}
\left\Vert \mathfrak{T}_{n,m}^{\pi }\right\Vert _{p}\overset{\mathrm{P}}{%
\underset{\pi }{\rightsquigarrow }}\Vert \mathbb{C}_{0}\Vert _{p}=\mathbb{T}%
^{(p)}\text{.}
\end{equation*}
By Lemma 10.11 in Kosorok (2008), it follows that
\begin{equation*}
P(\left\Vert \mathfrak{T}_{n,m}^{\pi }\right\Vert _{p}\leq c|\hat{Z})%
\xrightarrow{p}P(\mathbb{T}^{(p)}\leq c).
\end{equation*}%

In what follows, we introduce our Lemma A.2 and Lemma A.3, which formalize the continuous mapping theorem and the
conditional delta method for the conditional convergence. Lemma A.2 is a version of Theorem 10.8 of Kosorok (2008) and Lemma A.3 is obtained by extending the proof of Theorem 12.1 of Kosorok (2008) to the case where we have two independence copies of $\mathbb{X}_{1}$ and $\mathbb{X}_{2}$ in the limit, allowing that the laws of $\mathbb{X}_{1}$ and $\mathbb{X}_{2}$ to be different. See also Beare and Seo (2017).

\bigskip
\textit{\textbf{Lemma A.2. }}(Continuous Mapping Theorem) Let $\mathbf{A}$ and $%
\mathbf{B}$ be Banach spaces. Define a continuous map $f :\mathbf{A}%
\rightarrow \mathbf{B}$ at all points in a closed set $\mathbf{A}_{0}\subset
\mathbf{A}$. If $\xi _{N}^{\pi }\overset{\mathrm{P}}{\underset{\pi }{%
\rightsquigarrow }}\xi $ in $\mathbf{A}$, where $\xi $ is tight and
concentrates on $\mathbf{A}_{0}$, then $f (\xi _{N}^{\pi })\overset{%
\mathrm{P}}{\underset{\pi }{\rightsquigarrow }}f (\xi )$ in $\mathbf{B}$.

\bigskip
\textit{\textbf{Lemma A.3. }}(Conditional Delta Method) Let $\mathbf{A}$ and $\mathbf{%
B}$ be Banach spaces and let $f :\mathbf{A}_{f }\subset \mathbf{A}%
\rightarrow \mathbf{B}$ be Hadamard differentiable at $\mu \in \mathbf{A}%
_{f}$ tangentially to $\mathbf{A}_{0}\subset \mathbf{A}$, with
derivative $f _{\mu }^{\prime }$. Further let $\eta (N)\,$be a rate of
convergence and $\xi _{N}\in \mathbf{A}_{f }$ be an element which depends
on the data but not $\pi $. If $\eta (N)(\xi _{N}-\mu )\rightsquigarrow \mathbb{X}_{1}$ and $\eta(N)(\xi _{N}^{\pi }-\xi _{N})\overset{\mathrm{P}}{\underset{\pi }{%
\rightsquigarrow }}\mathbb{X}_{2}$ in $\mathbf{A}$
with tight limits $\mathbb{X}_{1}$ and $\mathbb{X}_{2}$ in $\mathbf{A}_{0}$,
we have $\eta (N)(f (\xi _{N}^{\pi })-f (\xi _{N}))\overset{\mathrm{P}}{%
\underset{\pi }{\rightsquigarrow }}f _{\mu }^{\prime }(\mathbb{X}_{2})%
\text{ in }\mathbf{B}$.

\bigskip \textbf{Proof of Lemma 3.4.} Note that for $\mathbf{u}=(u_{1},...,u_{d})\in
\lbrack 0,1]^{d}$ and each $\pi\in \mathbf{G}_{N}$, we have
\begin{eqnarray*}
\tilde{T}_{n,m}(\hat{Z}^{\pi })(\mathbf{u}) &=&\sqrt{\frac{nm}{n+m}}\left(
\frac{1}{n}\sum_{i=1}^{n}1(\hat{Z}^{\pi (i)}\leq \mathbf{u})-\frac{1}{m}%
\sum_{i=n+1}^{N}1(\hat{Z}^{\pi (i)}\leq \mathbf{u})\right)  \\
&=&\sqrt{\frac{nm}{n+m}}\left( \frac{(n+m)}{nm}\sum_{i=1}^{n}1(\hat{Z}^{\pi
(i)}\leq \mathbf{u})-\frac{1}{m}\sum_{i=1}^{N}1(\hat{Z}^{i}\leq \mathbf{u}%
)\right)  \\
&=&\sqrt{\frac{n(n+m)}{m}}\left[ \tilde{C}_{1,n}^{\pi }(\mathbf{u})-\left\{
\left( \frac{n}{n+m}\right) \hat{C}_{1,n}(\mathbf{u})+\left( \frac{m}{n+m}%
\right) \hat{C}_{2,m}(\mathbf{u})\right\} \right] \text{. \ \ }
\end{eqnarray*}%
Our condition on the rate of $\lambda _{n,m}$ implies $\lambda
_{n,m}-\lambda =o(n^{-1/2})$. Therefore, from the last equation we obtain
\begin{equation*}
\tilde{T}_{n,m}(\hat{Z}^{\pi })=\sqrt{n}(1-\lambda _{n,m})^{-1/2}\left\{
\tilde{C}_{1,n}^{\pi }-(\lambda \hat{C}_{1,n}+\left( 1-\lambda \right) \hat{C%
}_{2,m})\right\} +o_{p}(1)\text{.}
\end{equation*}%
In the same way, we can also verify that
\begin{equation*}
\tilde{T}_{n,m}(\hat{Z}^{\pi })=\sqrt{m}(\lambda _{n,m})^{-1/2}\left\{
\left( \lambda \hat{C}_{1,n}+\left( 1-\lambda \right) \hat{C}_{2,m}\right) -%
\tilde{C}_{2,m}^{\pi }\right\} +o_{p}(1)\text{.}
\end{equation*}%
employing the condition \bigskip $\lambda _{n,m}-\lambda =o(m^{-1/2})$.

In view of the conditional convergence, let $\pi$ be a random permutation uniform on $\mathbf{G}_{N}$, and let $\Psi _{n,m}^{1,\pi }$ and $\Psi _{n,m}^{2,\pi }$ be rescaled versions of the non-decaying terms,
\begin{eqnarray*}
\Psi _{n,m}^{1,\pi } &=&\sqrt{\frac{nm}{n+m}}\left\{ \tilde{C}_{1,n}^{\pi
}-\left( \lambda \hat{C}_{1,n}+\left( 1-\lambda \right) \hat{C}%
_{2,m})\right) \right\} \text{ and} \\
\Psi _{n,m}^{2,\pi } &=&\sqrt{\frac{nm}{n+m}}\left\{ \left( \lambda \hat{C}%
_{1,n}+\left( 1-\lambda \right) \hat{C}_{2,m}\right) -\tilde{C}_{2,m}^{\pi
}\right\} \text{. }
\end{eqnarray*}%
The conditional convergence of $\Psi _{n,m}^{1,\pi }$ and $\Psi _{n,m}^{2,\pi }$ can be obtained by Lemma 3.3 as,
\begin{equation}
\left(
\begin{array}{c}
(1-\lambda _{n,m})\tilde{T}_{n,m}(\hat{Z}^{\pi }) \\
\lambda _{n,m}\tilde{T}_{n,m}(\hat{Z}^{\pi })%
\end{array}%
\right) =\left(
\begin{array}{c}
\Psi _{n,m}^{1,\pi }+r_{n,m}^{(1)} \\
\Psi _{n,m}^{2,\pi }+r_{n,m}^{(2)}%
\end{array}%
\right) \text{ }\overset{\mathrm{P}}{\underset{\pi }{\rightsquigarrow }}%
\left(
\begin{array}{c}
(1-\lambda )\mathbb{\tilde{T}} \\
\lambda \mathbb{\tilde{T}}%
\end{array}%
\right)   \tag{14}
\end{equation}
in which both $r_{n,m}^{(1)}$ and $r_{n,m}^{(2)}$ are $o_{p}(1)$, and asymptotically negligible. 

We finish our proof by applying Lemma A.2 and Lemma A.3 to the result in (14). Note that the two following two conditions (i) and (ii) are satisfied for an application of the conditional functional delta method,
\begin{eqnarray*}
\text{(i) }\Psi _{n,m}^{1,\pi }\overset{\mathrm{P}}{\underset{\pi }{%
\rightsquigarrow }}(1-\lambda )\mathbb{\tilde{T}}\text{ \ and \ }
\Psi _{n,m}^{2,\pi }\overset{\mathrm{P}}{\underset{\pi }{%
\rightsquigarrow }}\lambda \mathbb{\tilde{T}} \text{ \ \ \ \ \ \ \ \ \ \ \ \ \ \ \ \ \ \ \ \ \ \ \ \ } \\
\text{(ii) }\sqrt{\frac{nm}{n+m}}\left( \lambda \hat{C}_{1,n}+\left(
1-\lambda \right) \hat{C}_{2,m}-\left( \lambda C_{1}+\left( 1-\lambda
\right) C_{2}\right) \right)  &\rightsquigarrow &\lambda \sqrt{1-\lambda }%
\mathbb{C}_{1}-(1-\lambda )\sqrt{\lambda }\mathbb{C}_{2},
\end{eqnarray*}
and we obtain
\begin{equation*}
\left(
\begin{array}{c}
\sqrt{\frac{nm}{n+m}}\left\{ \Phi (\tilde{C}_{1,n}^{\pi })-\Phi \left(
\lambda \hat{C}_{1,n}+\left( 1-\lambda \right) \hat{C}_{2,m})\right)
\right\}  \\
\text{\ }\sqrt{\frac{nm}{n+m}}\left\{ \Phi \left( \lambda \hat{C}%
_{1,n}+\left( 1-\lambda \right) \hat{C}_{2,m}\right) -\Phi (\tilde{C}%
_{2,m}^{\pi })\right\}
\end{array}%
\right) \text{ }\overset{\mathrm{P}}{\underset{\pi }{\rightsquigarrow }}%
\left(
\begin{array}{c}
(1-\lambda )\Phi _{\lambda C_{1}+(1-\lambda )C_{2}}^{\prime }\mathbb{\tilde{T%
}} \\
\lambda \Phi _{\lambda C_{1}+(1-\lambda )C_{2}}^{\prime }\mathbb{\tilde{T}}%
\end{array}%
\right) \text{.}
\end{equation*}
By the linearity of Hadamard derivatives we have
\begin{equation*}
\sqrt{\frac{nm}{n+m}}\left( \Phi (\tilde{C}_{1,n}^{\pi })-\Phi (\tilde{C}%
_{2,m}^{\pi })\right) \overset{\mathrm{P}}{\underset{\pi }{\rightsquigarrow }%
}\Phi _{\lambda C_{1}+(1-\lambda )C_{2}}^{\prime }\mathbb{\tilde{T}}\text{.}
\end{equation*}

\bigskip \textbf{Proof of Theorem 3.2.} Lemma A.1 and Lemma 3.4 imply that
under the null hypothesis, we have%
\begin{equation*}
\sqrt{\frac{nm}{n+m}}\left( \Phi (\tilde{C}_{1,n}^{\pi })-\Phi (\tilde{C}%
_{2,m}^{\pi })\right) \overset{\mathrm{P}}{\underset{\pi }{\rightsquigarrow }%
}\Phi _{\lambda C_{1}+(1-\lambda )C_{2}}^{\prime }\mathbb{\tilde{T}}_{0}%
\text{.}
\end{equation*}%
By using the same techniques as in Lemma 4.2 and Lemma 4.3 of Beare and Seo
(2017), we can verify that $\Phi (\tilde{C}_{1,n}^{\pi })$ differs from $%
\hat{C}_{1,n}^{\pi }$ by no more than $2n^{-1}$ and $\Phi (\tilde{C}%
_{2,m}^{\pi })$ differs from $\hat{C}_{2,m}^{\pi }$ by no more than $2m^{-1}$%
. This concludes,
\begin{equation*}
\sqrt{\frac{nm}{n+m}}(\hat{C}_{1,n}^{\pi }-\hat{C}_{2,m}^{\pi })\overset{%
\mathrm{P}}{\underset{\pi }{\rightsquigarrow }}\Phi _{\lambda
C_{1}+(1-\lambda )C_{2}}^{\prime }\mathbb{\tilde{T}}_{0}\text{.}
\end{equation*}%
Now applying our Lemma A.2 and\ Lemma 10.11 in Kosorok (2008), we have%
\begin{equation*}
P(\left\Vert \mathfrak{R}_{n,m}^{\pi }\right\Vert _{p}\leq c|\hat{Z})%
\xrightarrow{p}P(\mathbb{T}^{(p)}\leq c)
\end{equation*}%
for any continuity points $c\in (0,\infty )$. Therefore, the result (i) in
Theorem 3.2 follows. Next, suppose that the null hypothesis is false. Then
by Lemma 3.4 and Lemma A.2, the permutation distribution of $\left\Vert
\mathfrak{R}_{n,m}^{\pi }\right\Vert _{p}$ converges to $\Vert \Phi
_{\lambda C_{1}+(1-\lambda )C_{2}}^{\prime }\mathbb{\tilde{T}}\Vert
_{p}$, and the result (ii) in Theorem 3.2 follows from the divergence of $%
T_{n,m}^{(p)}$.

\pagebreak
	
\doublespace

\end{document}